\documentclass[12pt, prd,nofootinbib,longbibliography]{revtex4-2}
\usepackage{amsmath,amssymb,amsfonts,graphicx}

\begin{document}
\title{An asymptotic bootstrap method and its applications to Hermitian matrix models}
\author{David Berenstein$^{\dagger,\ddagger,\triangle}$ and Paula Garc\'\i a Mart\'\i nez $^{\dagger, \square}$}
\affiliation{$^\dagger$Department of Physics, University of California, Santa Barbara, CA 93106, USA}
\affiliation{$^\ddagger$Institute of Physics, University of Amsterdam, Science Park 904, PO Box 94485, 1090 GL Amsterdam,
The Netherlands}
\affiliation{$^\triangle$Delta Institute for Theoretical Physics, Science Park 904, PO Box 94485, 1090 GL Amsterdam, The
Netherlands.}
\affiliation{$^
\square$ Universidad Complutense de Madrid, Plaza de las Ciencias 1, 28040 Madrid, Spain
}

\begin{abstract} We propose an asymptotic bootstrap method to evaluate moment integrals that arise in problems related to hermitian matrix models. These are normalized integrals of the form $a_n=\int^{\infty}_{-\infty} x^n \exp(-V(x)) dx$ where $V(x)$ is a polynomial of $x$.  The method is applicable even  when the coupling constants in $V(x)= x^{2\ell}/(2\ell) + g_{2\ell-2} x^{2\ell-2}/(2\ell-2) + \dots$ are complex. We prove that the method converges asymptotically exponentially fast on a cone region determined by the first subleading coupling $g_{2\ell-2}$, which must have a positive real part and an absolute value of the argument less than $\pi/{\ell}$. We use our method to study the phase structure of Hermitian matrix models by constructing the orthogonal polynomials associated to these measures.
\end{abstract}

\maketitle

\section{Introduction}

The bootstrap philosophy to solve problems in mathematical physics, roughly stated, is that solving all the  consistency conditions of the problem of interest gives a (unique) solution to said problem.

In a quantum field theory, such conditions would be applied to the S-matrix for example, where one requires analyticity, unitarity and crossing symmetry. From our modern understanding of renormalizable quantum field theories, we know that there might be many such solutions parametrized by the coupling constants of the theory. Hence, the idea of finding a unique solution to the S-matrix program is not realistic on its own and more information needs to be given.

One can  look to specialize the problem by adding more symmetry for example. In two dimensions, if one chooses a critical point characterized by a conformal field theory, where there is an infinite dimensional symmetry, the goal is to obtain the data necessary to solve for the CFT correlators: conformal dimensions and OPE coefficients.  This program was successfully carried out in \cite{Belavin:1984vu} where minimal models were discovered and solved. Here there was a condition related to unitarity that constrained the spectrum of the theory, which coupled with the condition of the central charge $c<1$ was powerful enough to solve the models.

For higher dimensions, the conformal symmetry is finite and the techniques that are required to solve the problems are numerical.  The numerical Conformal Bootstrap has been very successful in bounding and finding the critical exponents of interesting field theories \cite{El-Showk:2012cjh} (for a review of the program see \cite{Poland:2018epd}) and this has produced a lot of interest in the numerical methods arising from this setup. This ha ignited  the exploration of the bootstrap methods, loosely thought of as solving a theory by consistency.

What has been actually very hard to find are proofs that the numerical bootstrap method converges in any of these settings. Here, we speak of convergence as finding the exact values of some data of interest with a controlled error determined in advance. 
More importantly, one would hope to know in advance what is the cost of computing the next digit (or quantity) of a difficult problem. In that sense, getting an analytic estimate of the errors is the better question to ask.
This problem is very difficult to solve in general. 

Although proving convergence might seem like a very mathematical problem with little physics in it, in exploring techniques to do prove convergence one might actually stumble on new algorithms or new methods to solve the problem.  Such new methods can be applied to cases that could not be covered by the original formulation of the problem. 

In this paper we work on a small class of such problems that arise from simplifying the bootstrap ideas to an extremely  tractable scenario.  There, we will be able to carry out this program, with some caveats: our proof of convergence is restricted to some conditions on the coupling constants.
However, our method, when it converges, extends beyond the initial formulation and allows solving some problems with a {\em sign problem}. We then apply this method to studying Hermitian matrix models at finite $N$ and then for illustration purposes, we look for the large $N$ phase transitions expected to arise in such models.

One can think of this restriction with the following example. Consider Riemann's zeta function $\zeta(s)\equiv\sum_{n=1}^\infty n^{-s}$. The definition converges when $\Re e(s)>1$. Other theorems show that the Riemann zeta function is meromorphic in $s$ and admits a unique analytic continuation to the complex plane.
If we want to evaluate it, we would do so in the region of convergence by its definition and we could be happy about this outcome, even if we do not have access to the othe regions. A similar procedure would work with any Taylor series with a finite radius of convergence. We would not ignore the problem until we find a solution that works everywhere. The fact that we sometimes can compute only in some region is no reason to throw our arms in the air waiting for something better to come along that covers every possible case. Although this might be desirable, progress can be made if the restrictions still allow for non-trivial phenomena (they can detect some phase transitions for example).

The problem we will study is the numerical determination of the moments of a distribution
\begin{equation}
a_n = \int_{-\infty}^\infty dx x^n \exp(-V(x))\label{eq:problem}
\end{equation}
where $V(x)$ is some (real) polynomial of $x$. This results in a simplified problem that can be analyzed with similar  methods to the ones that show up in the quantum-mechanical bootstrap approach of \cite{Han:2020bkb} (this method can be thought of as using hypervirial relations \cite{hirschfelder1960classical} plus positivity to get to physical answers). These seem to converge quickly to the spectrum of a 1-D Hamiltonian in a variety of settings \cite{Berenstein:2021dyf,Bhattacharya:2021btd,Aikawa:2021eai, Berenstein:2021loy,Tchoumakov:2021mnh}. The numerical techniques use positivity in an essential way and this positivity can be optimized to produce a semidefinite programming algorithm (SDP) to solve the problems \cite{Berenstein:2022unr}. It's also possible to solve problems with Robin boundary conditions \cite{Berenstein:2022ygg} (this also allows to study scattering \cite{Berenstein:2023ppj}) and more general problems with a discrete spectrum on an interval \cite{Sword:2024gvv,Thong:2026zvt}.
All these works suggest that the bootstrap method converges quickly in an abstract sense. However, when one evaluates the solutions of such problems using a standard SDP solver, one finds that the method is computationally intensive and much slower than other approximation methods. 

The type of problem we study was highlighted in \cite{Berenstein:2025itw} as a toy problem of the bootstrap. It has the same ingredients that the quantum mechanical bootstrap needs. It uses positivity of the measure as 
a condition similar to unitarity, it has similar Ward identities that result in a recursion. The main simplifying assumption is that this simplified problem  avoids the fact that in the quantum mechanical bootstrap problem one also needs to numerically scan over a parameter $E$ (the possible energy of an eigenstate). It was argued in \cite{Berenstein:2025itw} that convergence should be quick because of two mathematical properties of a Hankel matrix mode of moments of the distribution that appears in the problem:  it is known that it is a poorly conditioned positive matrix and the smallest eigenvalue goes to zero very quickly as one increases the rank of the matrix. What that means is that the Hankel matrix $M$ one needs is on the edge of becoming negative, and therefore it is easy to spoil positivity. This should lead to a finely tuned problem. 
Such an argument is a heuristic reason for the fast convergence. In a further development, for similar problems associated to measures on the circle \cite{Berenstein:2026wky} an alternative algorithm that bypassed positivity was found by exploiting the asymnptotic properties of Fourier moments of the distribution. This method showed that the bootstrap method for such measures based on positivity 
converged asymptotically exponentially fast, because it was possible to tie the error bounds to the asymptotic growth of the recursion that one needs to solve. Basically, it became the first proof of exponentially fast convergence for bootstrap problems and moreover it came with a new understanding that could avoid the semidefinite solve step in the algorithm and replace it with linear algebra. In that way, positivity was not required any longer and it was now possible to solve problems with complex valued potentials (what one would call sign problems for highly oscillating integrals).

In this paper we apply the same philosophy to the problem of measures on the line:  understanding the asymptotic behavior coupled with a weak version of positivity. 
We then show that we can get a  similar result 
for the problem of moments \eqref{eq:problem}, but which only converges when some additional conditions are imposed on the coupling constant of the potential. Again, we are able to sidestep the positivity step and replace it by linear algebra, resulting in a similar final result: we are able to extend the method to situations where the original formulation based on positivity does not apply.
These additional conditions do allow to extend the problem to complex measures with a sign problem, but they  converge only on a simple region of the complex plane of the leading coupling constant.

The paper is organized as follows. First, in section \ref{sec:positivity} we describe the bootstrap program based on positivity for the kind of moment problems that we study. The main ingredients are positivity and a recursion relation that arises from integration by parts identities. It is the recursion relation that will play a leading role in this paper.
In section \ref{sec:asym} we introduce our asymptotic bootstrap method, first in the simplest example of the moments associated to the potential $V(x)= x^4/4+g_2 x^2/2$. We show that for large moments, the ratios between consecutive moments are good variables that satisfy a simpler recursion and we also show that simple inequalities arising from positivity already bound these ratios asymptotically. Then we use that information to backpropagate the information to the initial moments. When $g_2>0$ we show that it converges, whereas for $g_2<0$ our method does not seem to converge. We show how to extend this to the complex plan and we give a proof of convergence for $\Re e (g_2)>0$, even in cases that are not covered by positivity.  Once we have established the simplest such setup, we present the general asymptotic bootstrap method for even potentials and we give a proof that it converges asymptotically exponentially fast when the leading non-trivial coefficient of  a polynomial $V(x)$ of degree $2\ell$, $g_{2\ell-2}$ sits inside a wedge with positive real part and characterized by $|\arg(g_{2\ell-2}|<\pi/{\ell}$. The proof of convergence relies on the norm of eigenvalues of a Jacobion being all in norm greater than one in the asymptotic regime. This condition is what specifies the wedge region of convergence. For the positive bootstrap, we show that the simple inequalities imply the form of the asymptotic ratio so that the proof applies to that method as well.
In section \ref{sec:applications} we apply our method to the study of hermitian matrix models by building the orthogoonal polynomials associated to the measure $d\mu \simeq dx \exp(-V(x))$. Because this problem is known to have instabilities due to poor conditioning of a positive matrix of moments, it requires many digits of precision to get good answers. This is what requires very high precision in intermediate steps. We show that we can track a phase transition where a one cut solution of the matrix model becomes a two cut solution.  Finally, in \ref{sec:concl} we conclude.

\section{Positivity}\label{sec:positivity}

In this paper we are considering the problem of determining numerically the moments of a positive measure on the real line, given as follows
\begin{equation}
a_n = {\cal N} \int_{-\infty}^\infty \exp(- V(x)) x^n dx
\end{equation}
where we normalize $a_0=1$, and $ V(x)$ is a polynomial of $x$ with real coefficients.
For convergence, the degree of $V$ needs to be even, $\deg(V)=2\ell$ and the coefficient of $x^{2\ell}$ in $V$ needs to be positive. A further simplification that can be used  is to ask that $V(x)=V(-x)$ is even itself, so that the odd $a_{2k+1}=0$ vanish identically by parity considerations. This simplification is not essential for the results in the paper. The constant ${\cal N}$ is a normalization constant. It can be 
absorbed into $V$ by a shift of $V$ by a constant.

Thought of as a measure,  $d\mu= \exp(-V(x))dx$ is a positive measure everywhere on $x$, meaning $\int_a^b d\mu>0$ for any interval $x\in (a,b)$, with $a<b$,  so it can be understood as a probability density.

Given this information, the following quantities are positive for any  polynomial $P(x)$ with complex coefficients
\begin{equation}
\int d\mu P(x)^* P(x) \geq 0 \label{eq:positivity}
\end{equation}
Expanding $P(x)= \sum_{j=0}^k s_j x^j$, we find that equation \eqref{eq:positivity} evaluates to a quadratic form in the $s_j$ coefficients given as follows
\begin{equation}
\int d\mu P(x)^* P(x)= \sum_{j,j'\leq k} s_j^* s_{j'} a_{j+j'}
\end{equation}
where we can extract a Hankel matrix with coefficients
\begin{equation}
M_{j,j'}= a_{j+j'}
\end{equation}
or equivalently, written in a notation where $a_n=\langle x^n \rangle$, we have that $M$ is a matrix of expectation values
\begin{equation}
    M= \left\langle\begin{pmatrix}
        x^0 & x^1 & x^2 &\dots\\
        x^1 & x^2 & x^3 &\dots\\
        x^2 &x^3 &x^4 &\ddots\\
        \vdots & \vdots &\ddots&\ddots
    \end{pmatrix}\right\rangle\label{eq:Hankel_mat}
\end{equation}

Since $\vec v^\dagger M\vec v\geq 0$ is positive for any vector $\vec v$ with coefficients $(\vec v)_j= s_j$, we find that the matrix $M$ is positive. This is written as follows $M\succeq 0$. The idea of the {\em positivity bootstrap }  is to use the inequalities associated to the positivity of $M$ to solve for the expectation values.

This is done with the aid of the following integration by parts identity
\begin{equation}
    \int_{-\infty} ^\infty \partial_x (x^{j} \exp(-V(x))=0 \label{eq:recursion}
\end{equation}
Let us expand $V(x)= \sum_{j=0}^{2\ell} \frac{g_j}{j} x^j$. Without loss of generality (by rescaling $x$), we can choose $g_{2\ell}=1$, so that the 
integration by parts identity is given by
\begin{equation}
j a_{j-1}- \langle x^j V'(x)\rangle=0 .
\end{equation}
This last expression can be written as
\begin{equation}
a_{j+2\ell-1}= ja_{j-1}- \sum_{k=0}^{2\ell-1} g_k a_{j+k-1}
\end{equation}
which shows that after $a_{0,\dots 2\ell-2}$ are known, the rest of the $a_{k\geq 2\ell-1}$ are determined recursively. For even potentials, we can discard all $a_{2k+1}$, so the recursion has fewer unknowns. 

The bootstrap method is to solve for the $a_{1, \dots,2\ell-1} $ as follows:
given a fixed degree $\tilde k$ of $P(x)$ above,  we find mathematically precise bounds 
\begin{eqnarray}
a^{(\tilde k)}_s \geq \min(a_s| M_{\tilde k}\succeq 0)\\
a^{(\tilde k)}_s \leq \max(a_s| M_{\tilde k}\succeq 0)
\end{eqnarray}
where $M_{\tilde k}$  the principal submatrix  of M, of  size  $(\tilde k+1)\times (\tilde k+1)$, where the recursion has been used to write it in terms of only $a_0=1, a_{1, \dots,2\ell-1} $. This submatrix is positive also, which is why we restrict the minimization, maximization to this subset. This is expected to converge to the correct answer when $\tilde k\to \infty$.

Each of the problems characterizing the maximum and minimum are semidefinite programming problems ( a subset of convex optimization problems), and when we increase $\tilde k$, the range of the allowed intervals decrease. It has been argued that as $\tilde k\to\infty$, the convergence is fast, because in the measures under consideration, the smallest eigenvalue of $M_{\tilde k}$ goes to zero faster than 
polynomially. What this means is that it is hard to find values of the unknown $a_s$ that are compatible with the constraint.

Since the measure exists and decays sufficiently fast, the unknown data must have a solution. This is in contrast to the quantum mechanical problem \cite{Han:2020bkb}, where we assume that we have an eigenstate of energy $E$ and then proceed to check if a positive measure exists, which is matched to the quantum mechanical measure $d\mu \simeq |\psi(x)|^2$. The existence of a measure is controlled by positivity and a particular behavior in the asymptotics (see for example \cite{schmudgen2017moment}), so one has to check positivity.

If we know ahead of time that the answer converges when $\tilde k\to \infty$, instead of finding bounds, we can find one interior point of the convex set of allowed solutions for each $\tilde k$. This is done by optimizing over the unknowns, maximizing the minimal eigenvalue of $ M_{\tilde k}$. This is done by considering the auxiliary problem
\begin{equation}
\lambda^*= \max(\lambda| M_{\tilde k}-\lambda {\bf 1}\succeq 0)
\end{equation}
which is another convex optimization problem of the same kind. This is trying to maximize the positivity of  $M_{\tilde k}$, by making the smallest eigenvalue of $M$ (which is equal to $\lambda^*$ in the solution) to be as large as possible \cite{Berenstein:2022unr}. 

The solution of these problems produces both $\lambda^*$, and the optimal value of the matrix $ M^*_{\tilde k}(a_s)$ (the interior point). We can solve for the $a_s^*$ by reading the components of $M^*$. So this provides a direct estimate of the solution for each $\tilde k$. Since a positive solution of this problem exists for any 
$\tilde k$ (the true solution always satisfies this condition), this optimization problem will give an approximation to the correct values of the unknowns, although we lose the information of the mathematical bounds
at each $\tilde k$. Numerically, this is a more efficient path to the solution, as we only need to solve on problem for each $\tilde k$.
If we believe that this procedure converges exponentially fast, we can use the difference between consecutive values of $(a^{\tilde k})^*$ to estimate the errors of the approximation.

The main problem in the above setup is practical: solving these optimization problems is numerically expensive at large $\tilde k$. Coupled with the problem that the $a_n$ grow factorially and that the smallest eigenvalue of $M_{\tilde k}$ is exponentially small, we find situations where solving the problem produces a poorly conditioned matrix $M^*_{\tilde k}$ and we need arbitrary precision arithmetic to find the solution. 
This poor conditioning is also the reason that the solution converges fast in the first place, so it is not a feature that can be easily fixed.

The goal of this paper is to get around these limitations by exploiting additional information that can be extracted asymptotically. 
Essentially, the numerical problem given to an SDP solver doesn't know intrinsically that the $a_n$ grow factorially and the matrix $M_{\tilde k}$ is poorly conditioned. The idea is to find an alternative method that exploits this additional asymptotic information to obtain another estimate for  the $a_n$ that bypasses the SDP step all together. We can then check positivity of the solution (up to some value $\tilde k$) we found if needed. We will study these asymptotic properties in the next section and show how these can be used to get around positivity in some cases, so that one can study some problems where the $g_s$ coefficients are complex.

\section{Asymptotic analysis}\label{sec:asym}

In this section and in the rest of the paper, for simplicity, we will restrict ourselves to situations where the potential is even $V(x)=V(-x)$.
In that case, we will make the substitution $b_s= a_{2s}$ and work with the recursion for the $b$.

Let us start with a simple observation. The $b_s$ grow factorially. This can be seen as follows
\begin{equation}
b_s= {\cal N} \int_{-\infty}^\infty \exp(- V(x)) x^{2s} dx
\end{equation}
We are interested in the growth at large $s$. We can estimate the integral by a saddle point method, where we consider the saddle $x_s$ that solves the potential problem
$\partial_x (2s\log(x) -V(x))=0$. This results in solving for a root of 
\begin{equation}
    V'(x)\simeq 2s/x
\end{equation}
Since $2s$ is a large number, $x$ grows with $s$ and we can simplify the problem by keeping only the leading terms of $V(x)$. We find this way that approximately 
\begin{equation}
x_s^{2\ell} \simeq 2s
\end{equation}
and that $b_s\simeq \exp( -x_s^{2\ell}/(2\ell) +2s \log(x_{s}))$.
This gives us an estimate
\begin{equation}
b_s \simeq \exp\left[ -(2s)/{2\ell}+\log(2s) 2s/{(2\ell)} \right]
\simeq \Gamma[ s/\ell] \exp(2s \log(2\ell)/(2\ell))
\end{equation}
where we are ignoring subleading terms in the exponential and the determinant of the saddle. This shows that the moments $b_s$ grow very fast. They are also obviously positive.

For similar problems in the case of integrals on the circle, where the recursive moments $a_s$ are Fourier modes, rather than $b_S$, one can argue that these asymptotically go to zero. That asymptotic property was used to solve the problem by approximating $a_s= 0$ at sufficiently large $s$ and solving backwards using linear algebra to find the finite number of unknowns
$ a^*_{1, \dots}$. It is clear that we cannot do this in this case, as the $b_s$ necessarily grow, and they grow quickly.

Consider now the minor of $M$ given by
\begin{equation}
   M_{s, small}= \begin{pmatrix}
        b_{s-1} & b_{s}\\
        b_{s} & b_{s+1}
    \end{pmatrix} \succeq 0
\end{equation}
which is positive when we have evaluated the moments. This is a subset of the positivity constraints $M$. How can we use this positivity to make inroads? It is convenient to make a change to variables that grow more slowly, given by ratios
\begin{equation}
r_s= \frac{b_{s+1}}{b_s}\geq 0.
\end{equation}
Evaluating the determinant of $M_{s, small}$ and after some simple manipulations, we find that
\begin{equation}
    r_{s}\geq r_{s-1} \geq 0 \label{eq:mono}.
\end{equation}
That is, we get a monotonically increasing sequence of ratios $r$. Our goal is to leverage these inequalities to make progress in proving convergence of the method. As we will see later in this paper, this simplifying procedure only works for some values of the coupling constants $g_k$. 
One can think of this step of analyzing asymptotic behavior in large $s$ as being similar in spirit to analytic approaches to the conformal bootstrap based on large spin asymptotics \cite{Fitzpatrick:2012yx} (see also \cite{Hartman:2022zik}).

\subsection{ A warm up problem: the computation of a Bessel ratio  }

It is convenient at this point to start with an example, where we take the simplest potential where the recursion has some unknown quantity to determine. This is given by the quartic potential
\begin{equation}
V(x) = \frac {g_2} 2 x^2+\frac {x^4} 4.
\end{equation}
This has the advantage of being soluble analytically in terms of Bessel functions.
For example, the partition function integral can be computed
\begin{equation}
    Z[g_2]=\int_{-\infty}^\infty\exp\left[-\frac {g_2 x^2}{2}-\frac {x^4}4\right]=
    \sqrt{\frac{g_2}{2}} \exp\left(-\frac{g^2_2}{8}\right)K_{1/4}\left[\frac{g_2^2}{8}\right].
\end{equation}
which gives us the normalization constant ${\cal N}=1/Z[g_2] $. 
Using $b_1=a_2=\langle x^2\rangle =-2\partial_{g_2}\log( Z[g_2])$ we find that 
\begin{equation}
    b_1= -\frac{1}{2} g_2
   \left(1-\frac{K_{\frac{3}{4}}\left(\frac{g_2^2}{8}\right)}{K_{\frac{1}{4}}\left(\frac{g_2^
   2}{8}\right)}\right),\label{eq:Besselratio}
\end{equation}
which requires computing a ratio of Bessel K functions. These can be evaluated to arbitrary precision if we want to. The formula written this way is valid so long as $\Re e(g_2)>0$. With this information we can compare the solution we will obtain by the asymptotic method we are trying to establish to an exact result. This way we can test the method, rather than solve a problem with no known solution.

To do that,  first we evaluate the  recursion for the even moments $b$,
\begin{equation}
    b_{s+2}= (2s+1)b_s - g_2 b_{s+1},
\end{equation}
which coupled with $b_0=1$ shows that the only unknown is $b_1$. Dividing by $b_{s+1}$, we find a similar recursion for the $r_s$ ratio sequence, giving us
\begin{equation}
    r_{s+1} = (2s+1) \frac 1{r_s} -g_2.\label{eq:recursionr}
\end{equation}
Now we can use the monotonicity inequalities \eqref{eq:mono} inside the recursion relation to find the following two inequalities (evaluating the recursion at $s,s-1$ respectively):
\begin{eqnarray}
(2s+1) \frac 1{r_s} -g_2\geq r_s\\
(2s-1) \frac 1{r_{s}} -g_2\leq r_{s}.
\end{eqnarray}
This is progress, as $r_s$ cannot be too large or too small. Indeed, $r_s$
is nested between the two positive roots $\chi_{\pm}$ of the two polynomials
\begin{eqnarray}
-(2s+1) +g_2 \chi_++\chi_+^2&=&0 \\
-(2s-1) +g_2 \chi_-+\chi_-^2&=&0
\end{eqnarray}
We find that 
\begin{eqnarray}
    \chi_+&=& \frac{-g_2+\sqrt{g_2^2+4(2s+1)}}2\simeq \sqrt{2s+1}-\frac {g_2}2\\
    \chi_-&=& \frac{-g_2+\sqrt{g_2^2+4(2s-1)}}2\simeq \sqrt{2s-1}-\frac {g_2}2.
\end{eqnarray}
These show that the difference between $\chi_{\pm}$ is small and of order $1/\sqrt{s}$, so the ratios $r$ are fairly constrained. A natural midpoint is to take 
\begin{equation}
    \chi_0\simeq \sqrt{2s}-g_2/2\label{eq:chi0}
\end{equation} for estimates. 
Indeed, we expect that $r_s$ admits an asymptotic series expansion in $1/\sqrt{s}$ that starts with $r_s\simeq \sqrt{2s}-\frac {g_2}2+O(1/\sqrt{s}) $, which arises from evaluating the ratio using saddle point methods. In that sense, the   $\chi_0$ approximation is optimal at this order in the asymptotic expansion.

Now, we can use $r_s= b_{s+1}/b_s \equiv \sqrt{2s}-\frac {g_2}2 =\chi_0$ to solve for $b_1$. We can instead also use $\chi_{\pm}$ if we want to.  Notice that both $b_{s+1}$ and $b_s$ are linear functions of $b_1$. Taking this ratio and setting it equal to the estimate of $r_s$ gives us a way to evaluate an approximation to $b_1$. This estimate $b^{(s)}_1$ is obtained by solving a linear algebra problem. This is our asymptotic estimate for $b_1$, which depends on the cutoff order $s$ (this is sometimes known as the level of the truncation in the SDP bootstrap setting). 
Notice that instead of the full set of inequalities from positivity, we have relaxed most of them and kept one such inequality to do the work.
This set of inequalities is implicitly contained in the estimate for $r_s$ at large $s$.

It is important to notice that this way of solving for $b_1$ is very similar in spirit to the asymptotic technique used in \cite{Berenstein:2026wky} (see also the analysis \cite{Huang:2025sua} for quantum mechanics on a periodic potential). In that case the problem was to solve for the Fourier coefficients of a positive measure  on the circle. We have been able to reduce the problem  to a linear algebra problem that can be done with high precision arithmetic, rather than solving the semi-definite programming problem directly. Here, the inequalities are the starting point, but by the end, we are solving a different type of problem: a linear algebra problem. 

Notice also that since we are now basing our solution on solving a linear algebra problem, we can extend the realm of applicability of our technique to cases that would not be covered by positivity. Namely, taking $g_2$ to be  a complex number.  In that case, the integrals one is trying to calculate are oscillatory (there is a sign problem) and certain numerical  methods to estimate the integrals (like Monte Carlo) would fail. 

Our next goal is to explore if and when this procedure produces the desired solution to the original problem. Moreover, if the method works, we want to estimate the rate of convergence of the answer (how well the method works).

\begin{figure}[ht]
\includegraphics[width=8cm]{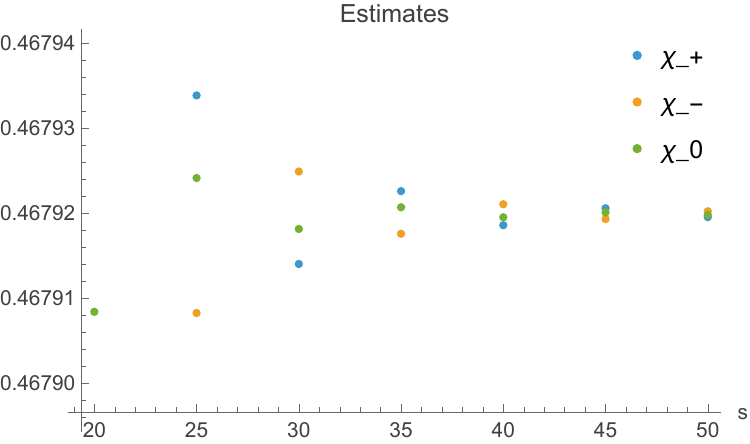}
    \caption{Asymptotic bootstrap method solution for Bessel ratio \eqref{eq:Besselratio} at $g_2=1$, with the three estimates from $\chi_{\pm}(s),\chi_0(s)$. We see convergence to the correct answer, $b_1=0.467919916973665188\dots.$}\label{fig:Bessel}
\end{figure}

\begin{figure}[ht]
\includegraphics[width=8cm]{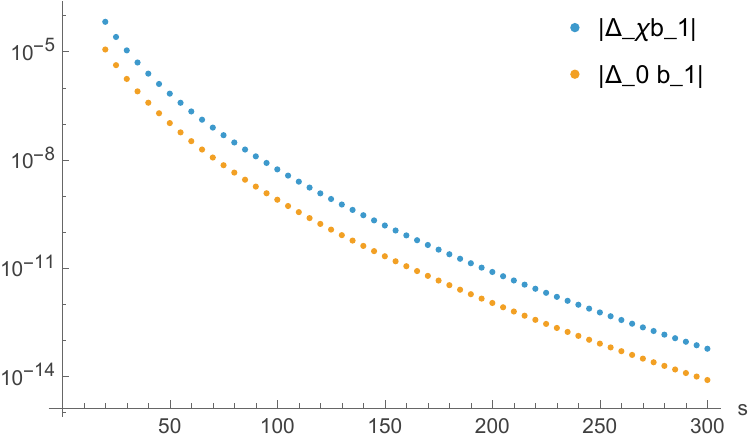}
    \caption{Intrinsic Errors $\Delta_\chi(b_1)=|b_1(\chi_+(s))-b_1(\chi_-(s))|$ and $|b_1(\chi_0(s))-\hbox{Exact}|$ for $g_2=1$, as a function of the recursion stopping point $s$. We see that the error decreases very quickly, so that at $s=300$ we can already evaluate the Bessel ratio \eqref{eq:Besselratio} to 14 decimal places.}\label{fig:ErrorBessel}
\end{figure}

Let us explain this method in more detail. It is convenient to organize the 
recursion as a multiplication of matrices as follows
\begin{equation}
   \vec v_{s}= \begin{pmatrix}b_s\\
    b_{s+1}\end{pmatrix}= \begin{pmatrix} 0 &1\\
    (2s-1)& -g_2 \end{pmatrix}\begin{pmatrix}b_{s-1}\\b_s\end{pmatrix}= M(s). \vec v_{s-1}
\end{equation}
Our goal is to compute the matrix $U(s)=M(s)\dots M(1)$, which is independent of the initial conditions. Then 
we have that $ r_s= v_{s,2}/v_{s,1}$, with initial conditions $v_{0,0}=1, v_{0,1}=b$. The quantities $v_{s,1},v_{s,1}$ are determined by multiplication by $U$ on the initial data. 
We solve for $b$ by assigning a value to $r_s$ for some $s$ and changing $s$. We use the three values $\chi_+,\chi_-, \chi_0=\sqrt{2s}-\frac{g_2}2$. The results are shown in figures \ref{fig:Bessel} and \ref{fig:ErrorBessel} which show fast convergence to the correct values when we choose $g_2=1$.

\begin{figure}[ht]]
\includegraphics[width=10cm]{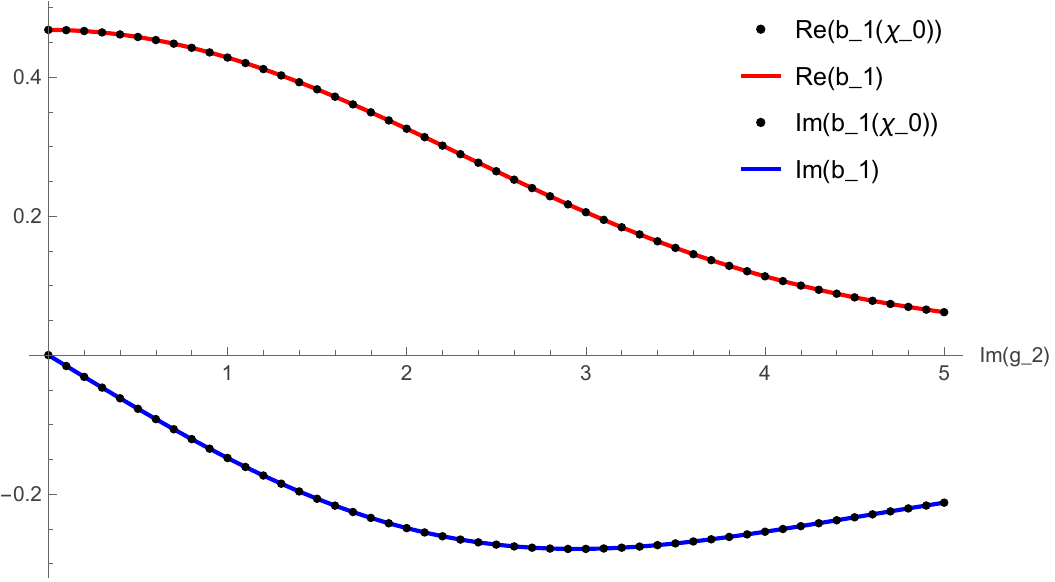}
    \caption{Evaluation of the Bessel ratio of equation \eqref{eq:Besselratio}. We use complex coupling constants $g_2=1+i w$, where $w\in[0,5]$. This is compared to the exact value. The cutoff is at recursion level $s=50$ for the matrices.   }\label{fig:Complexg2}
\end{figure}

We can  now also explore how well the code performs when we make $g_2=1+i s$
into a complex variable. The result of these computations is shown in figure \ref{fig:Complexg2} for a list of values of $s$ between $0$ and $5$.
Both the real and imaginary part are very close to the correct answer, so that the error is not visible in the graph.
Our results show that the code reproduces (converges to) the correct answer for complex values of $g_2$ also. In this sense, out asymptotic bootstrap has liberated us from the requirement of positivity.

\begin{figure}[ht]
\includegraphics[width=8cm]{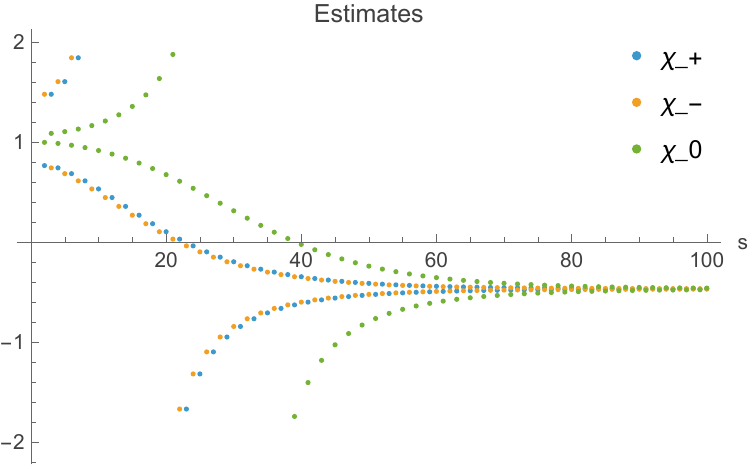}
    \caption{Asymptotic bootstrap method solution for Bessel ratio \eqref{eq:Besselratio} at $g_2=-1$, with the three estimates from $\chi_{\pm}(s),\chi_0(s)$. We see that it tries to converge to a negative value, whereas the correct value should be at $b_1=1.0418$ }\label{fig:Bessel2}
\end{figure}

However, not everything works. Consider now the case of negative coupling $g_2=-1$.
A similar plot to figure \ref{fig:Bessel2} for that case shows that the answer fails to converge. We also notice that something strange is going on, as $\chi_0$ seems to be outside the interval of solutions of $\chi_{\pm}$ at large $s$.
This can be explained if we have the intervals wrong: what is allowed at large $s$ is the region outside the band excluded by the bounds.
Essentially, at the beginning of the recursion the interval of allowed values seems to grow rather  than decrease. Moreover, it grows so much that it starts allowing negative values also. The ``convergence" we see is to an excluded value. 

The lesson that we learn from this exercise is that this asymptotic method we are developing sometimes works and sometimes does not. Our goal now is to understand when it actually works. We have stated two goals for this section. The first one is  to mathematically prove that the (positive) bootstrap method converges to the correct solution. The method we are developing is a subset of that program where we impose a minimal inequality to guide us to a solution, plus some additional asymptotic information.
Since it sometimes work, we can ask about convergence estimates in the appropriate region of the coupling constant $g_2$.
A second  goal is to extend the method beyond positivity, so that we can apply the solutions to situations with a sign problem. We have succeeded in this endeavor sometimes. Our goal is now to figure out in what region of the complex plane we get good results.

\subsection{Convergence region and estimates of convergence for the Bessel ratio}

We want to understand convergence in our toy problem. The idea is to estimate how the region of allowed $r$ increases or decreases under the recursion map \eqref{eq:recursionr}. We have
\begin{equation}
r_{s+1}=\frac{2s+1}{r_s}-g_2,
\end{equation}
so it follows that if we want to do error propagation, we need to compute the derivative
\begin{equation}
\frac{\partial r_{s+1}}{\partial r_s}= -\frac {2s+1}{r_s^2}
\end{equation}
Now we use $r_s\simeq \sqrt{2s}-g_2/2+O(1/\sqrt{s})$ to find that $r_s^2\simeq 2s-\sqrt{2s} g_2+O(1)$. This way we get that dividing the numerator and denominator by $2s$
\begin{equation}
\left|\frac{\partial r_{s+1}}{\partial r_s}\right|\simeq
\left|\frac{1}{1-g_2/\sqrt{2s}}\right|= 1 +\frac{\Re e [g_2]}{\sqrt{2s}}+O(1/s),\label{eq:Jac}
\end{equation}
where we have expanded $g_2$ into real and imaginary part, and we are also expanding at large $s$.
So the asymptotic map from $r_s\to r_{s+1}$ expands a small region in the complex plane if $\Re e(g_2)>0$. This means that if we know $r_{s+1}$ with some precision $\delta r_{s+1}$, via the inverse map we increase the precision of $r_s$ if the forward Jacobian is greater than one.
That can be interpreted as saying that we increase the information in $r_s$ (the correct number of digits of $r_s$).  On the other hand if the map is contracting, when  $\Re e (g_2)<0$, then the region of allowed $r_s$ increases in size and we are not gaining information.

Consider the chain rule to (back)propagate this result to the target value of $b_1\equiv r_0$,
\begin{equation}
   \left|\frac{\partial r_{s+1}}{\partial r_0}\right| =
   \prod_{k\leq s}\left|\frac{\partial r_{k+1}}{\partial r_k}\right|
   \simeq \exp \left[\sum_{k=1}^s \frac{\Re e [g_2]}{\sqrt{2k}}\right]
   \simeq \exp\left(\int_0^s \frac{\Re e [g_2]}{\sqrt{2k}}dk\right),
\end{equation}
so we estimate that the amplification factor of the region near the correct value of $r_0$ is given by 
\begin{equation}
      \left|\frac{\partial r_{s+1}}{\partial r_0}\right|\simeq 
      \exp\left( \Re e[g_2] \sqrt{2s} )\right) 
\end{equation}
Equivalently, when we think of precision of the answer,  
\begin{equation}
     | \delta r_0| \simeq \exp\left( -\Re e[g_2] \sqrt{2s} )\right) \delta r_{s}.
\end{equation}
We have that since $r_0=b_1/b_0=b_1$,  we find a precision of order
\begin{equation}
\delta b_1\simeq \exp\left( -\Re e[g_2] \sqrt{2s} )\right)/ {\sqrt{s}}
\end{equation}
The corrections on the integral start at  $\int dk/k\simeq \log(s)$, so these correct the power of $s$ appearing in $\delta b_1$. Beyond that, the result is convergent, so we find that
\begin{equation}
\delta b_1\simeq \exp\left( -\Re e[g_2] \sqrt{2s} +\alpha \log(s) +O(1))\right) 
\end{equation}
What this means is that the method converges faster than any polynomial bound on $s$. This is usually expressed as saying that the method converges asymptotically exponentially fast.

\begin{figure}[ht]
\includegraphics[width=8cm]{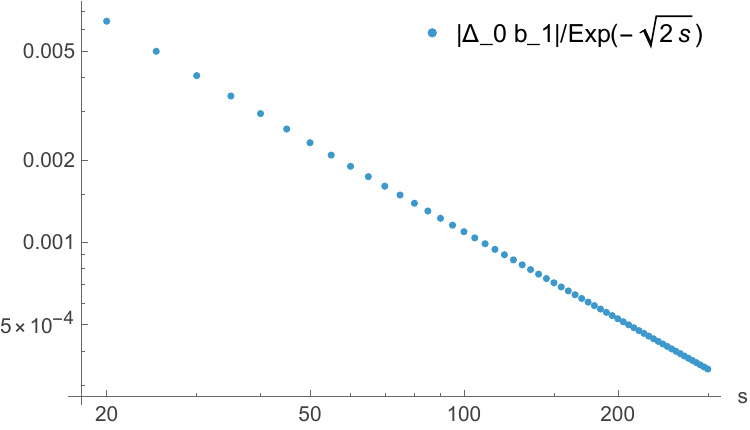}
    \caption{Asymptotic bootstrap method error relative to theoretical estimate, with the error coming from $|\chi_0(s)-b_1|$, where $b_1$ is the exact value. We divide by $\exp(-\Re[g_2]\sqrt{2s}$, which is the leading estimate of the error.  }\label{fig:BesselRelativeerror}
\end{figure}

In this case, the method converges in half of the complex plane. 
We see that our way of understanding convergence also explains the convergent results we got in the complex plane at $g_2=1+i \Im m (g_2)$. 
Notice that our results converge faster when $\Re e[g_2]$ is large.
This is also a regime where we would need to use the asymptotic expansion of Bessel functions for large argument if we want to compute the ratio exactly. Armed with this estimate, we can now plot the error, relative to the analytic estimate in figure \ref{fig:BesselRelativeerror}. We see that in the log-log plot we have a straight line by inspection, confirming our analytics.

The way we are supposed to interpret these results is as follows: the positive bootstrap method converges at least exponentially fast for $g_2>0$, since in that case the asymptotic method is a relaxation of the full set of inequalities arising from positivity. Moreover, notice that
we have also proved convergence when we extend the asymptotic method to some regions of the complex plane. With more effort, it is clear that the slope of the residual should be accessible to the asymptotic analysis.
The estimate can also be improved by using additional terms in the asymptotic expansion of $r_s$, so that an analog of $\chi_0$ is more precise ab initio. Understanding that problem, although interesting,  is beyond the scope of the present paper. We have already proved what we needed for our numerical method in the restricted context of this exactly solvable problem.

There is an additional point we would like to make here. Notice that the recursion relation \eqref{eq:recursionr} can also be solved as follows
\begin{equation}
r_s= \frac{2s+1}{g_2+r_{s+1}}
\end{equation}
so that the solution can be written as a continued fraction
\begin{equation}
    r_0= \frac{1}{g_2+r_1}=\frac{1}{g_2+\frac 3{g_2+ r_2}}=\frac{1}{g_2+\frac3{g_2+\frac 5{g_2+\dots}}}
\end{equation}
Such expressions are formally exact solutions of the Schwinger-Dyson equations of the model \cite{Fiol:2025gkq}. Notice that our approximation chooses a particular value for $r_{s}\simeq \sqrt{2s}-g_2/2$ at some large $s$. That is an important part of how we should evaluate such an expression.
Regardless, this way one can produce rational approximations to the ratios of moments as functions of $g_2$ (for more information, see \cite{Fiol:2025gkq} and references therein). We have proven convergence so long as $\Re e (g_2>0)$, which suggests that the formal continued fraction can be a non-perturbative definition in some region of the $g_2$ complex plane. In the work of Fiol et al. \cite{Fiol:2025gkq} it seems that a crucial feature to exploit is that some of these correlation functions are Stieltjes functions of $g_2$ and they also obtain convergence for the condition $g_2>0$.
Exploring these ideas further is beyond the scope of the present work.

\subsection{The general asymptotic bootstrap algorithm for even potentials}

We are finally ready to tackle the more general problem of solving for the moments of integrals of the type
\begin{equation}
  b_m={\cal N}  \int_{-\infty}^\infty x^{2m} \exp(-V(x)) dx
  \label{eq:moments}
\end{equation}
with 
\begin{equation}
    V(x) = \sum_{k=1}^\ell \frac{g_{2k}}{2k}x^{2k}
\end{equation}
with a normalization restriction for the leading term $g_{2\ell}=1$.
Our goal is now to generalize the formalism in the toy problem to the more general setting. 

The recursion relation for the $b_m$ is defined by $b_0=1$ and
\begin{equation}
    b_{s+\ell}= (2s+1)b_s-\sum_{k=1}^{\ell-1} g_{2k} b_{s+k} \label{eq:recb2}
\end{equation}
so we only need to determine $b_1,\dots, b_{\ell-1}$ to solve the problem. 
We now need a list of values, rather than a single value.

For convenience, we write the recursion also as follows
\begin{equation}
\vec v_s =\begin{pmatrix}
    b_s\\
    b_{s+1}\\
    \vdots\\
    b_{s+\ell-1}
\end{pmatrix}= \begin{pmatrix} 0&1&0&0&\dots\\
0&0&1 & 0&\dots\\
\ddots &\ddots &\ddots &\ddots &\ddots\\
0& 0& 0& \dots &1\\
(2s-1) & -g_{2} & -g_4 &\dots &-g_{2\ell-2}\end{pmatrix}
\begin{pmatrix}
    b_{s-1}\\
    b_{s}\\
    \vdots\\
    b_{s+\ell-2}
\end{pmatrix}=M(s)\vec{v}_{s-1} 
\end{equation}
Similarly, we introduce the ratios $r_s= b_{s+1}/b_s$. The recursion can be written in terms of only the $r_s$, by taking \eqref{eq:recb2} and dividing by $b_{s+\ell-1}$ to obtain
\begin{equation}
r_{s+\ell-1}= \frac{(2s+1)}{r_s\dots r_{s+\ell-2}}- g_{2\ell-2}-O(1/r_s)\label{eq:recr_long}
\end{equation}
where we are focusing on the two leading terms in $s$. Since the $r_s\dots r_{s+\ell-1}$ are increasing, we can find similar inequalities to those that determined $\chi_\pm$ and the midpoint $\chi_0$. A similar estimate here starts with all of them being the same, and then approximately
\begin{equation}
    r_s\simeq \frac{(2s+1)}{r_s^{\ell-1}}-g_{2\ell-2}
\end{equation}
so that we have
\begin{equation}
r_s^\ell+g_2 r_s^{\ell-1}-(2s+1)\simeq 0\label{eq:poly}
\end{equation}
This results in 
\begin{equation}
    r_s\simeq (2s)^{1/\ell}-g_{2\ell-2}/{\ell}+O(1/s^{1/\ell})\label{eq:leading_asym}
\end{equation} where we generically expect an asymptotic series in $s^{-1/\ell}$. Indeed, we should point that we are picking one of the solutions of the solutions of \eqref{eq:poly}, which is the one on the positive real axis when the polynomial has real coefficients.
At very large $s$ the other roots are very close to uniformly placed on a circle of radius $r_s$. These would be different saddles of the integral. Implicitly, we assume that this saddle dominates even when 
we move $g_{2\ell-2}$ in the complex plane. This issue and its possible connection to resurgence is beyond the scope of the present paper.

This series is also expected when we have complex coupling constants rather  than real ones.  If we want to be more precise, we keep all powers of $r_s$ in the polynomial we are trying to take a root of. 
It is clear that the term that varies is associated to the $2s+1$, so it results in corrections $\delta r_s\simeq O(1/s^{1-1/\ell})$. If we ignore the full polynomial, we get $\delta r_s\simeq O(1/s^{1/\ell})$ which will be good enough for our purposes. 

The algorithm we need to use is  to first take the matrix product $U(s)=M(s)\dots M(1) $, which is independent of the initial condition, and then we set $v_{s,m+1}/v_{s,m}=r_{s+m+1}\equiv 
\chi_{s+m}$ where $\chi_{s+m}= (2s)^{1/\ell}-g_{2\ell-2}/{\ell}$, 
where $\vec v_s= U(s) \vec v_0$. That is, we get a linear equation for each ratio of consecutive components of $\vec v_s$. This is exactly a linear problem with the same number of unknowns as the initial condition requires.
Hence, we can solve for values $b_1^{(s)}\dots b_{\ell-1}^{(s)}$ at each iteration order. 

We can also make refinements of the problem inequalities that will usually shift the subleading pieces in $\chi_s$, so that we can obtain analogs of $\chi_{\pm}$ for each $r_s$. Since we have multiple linear equations, we need to make such a choice for each of the $\chi_s$ appearing in the inequalities. This will produce a small (rectangular) region of the set described with parameters   $r_{s}\dots r_{s+\ell-1}$. 

If we have complex coupling constants instead,  we should think of having a small open set around the correct (complex) asymptotic solution to quantify the errors, but the idea of having exact positivity bounds is gone. The error for the $\delta r_s$ in this case is of the order of the terms that have been ignored in the asymptotic expansion of $r_s$. Regardless, the problem we need to solve is linear in the $v$, determined by the $r_s$, which are holomorphic in the complex coupling constants $g_{2k}$.

Notice that the procedure we have defined contains the same ingredients that were used in the toy model problem of the ratio of $K$ Bessel functions. Since in that case the answer converged only for a subset of the complex plane $g_2$, we can now ask what is the region of the complex parameters $g_2, \dots g_{2\ell-2}$ where the procedure converges. To study convergence, we need to generalize the problem of propagation of errors to multiple variables. In order to do that, we need to understand by how much the forward Jacobian matrix grows or shrinks the different directions. We will be able to guarantee convergence in the region where all directions grow in the asymptotic regime.

\subsubsection{Convergence}

We should begin from the recursion \eqref{eq:recr_long}
\begin{equation}
r_{s+\ell-1}= \frac{(2s+1)}{r_s\dots r_{s+\ell-2}}- g_{2\ell-2}-O(1/r_s)\label{eq:long2}
\end{equation}
Remember that we need to solve for $b_{1,\dots,\ell-1}$. Since we need to keep track $\ell-1$ unknowns, we need to keep track of the same number of unknowns in the $r_s$, namely  of $r_s, \dots, r_{s+\ell-2}$. We want to think of the recursion as a map from the set with coordinates
$(r_s, \dots, r_{s+\ell-2})$ to the set $(r_{s+1}, \dots, r_{s+\ell-1})$.
This map deletes the first coordinate, shuffles the last $s-2$ coordinates to the first $(s-2)$ coordinates of the new point, and computes $r_{s+\ell-1}$ from equation \eqref{eq:long2}. We will call this map $R^{(s)}:r^{(s)}\to r^{(s+1)}$ (in analogy with the matrix $M(s)$).  

The Jacobian $J(s)$ of $R^{(s)}$ is given by the matrix
\begin{equation}
    J(s) =\begin{pmatrix}  0&1&0&0&\dots\\
0&0&1 & 0&\dots\\
\ddots &\ddots &\ddots &\ddots &\ddots\\
0& 0& 0& \dots &1\\
\frac{\partial r_{s+\ell-1}}{\partial r_s} & \frac{\partial r_{s+\ell-1}}{\partial r_{s+1}} & \frac{\partial r_{s+\ell-1}}{\partial r_{s+2}}&\dots& 
\frac{\partial r_{s+\ell-1}}{\partial r_{s+\ell-2}}
    \end{pmatrix}
\end{equation}
By the chain rule, the total Jacobian from the initial $r$ to the final $r$ is given by the product of the matrices $J_{tot}= J(s)
\dots J(0)$.
Now let us put the asymptotic behavior of $r_s$ from \eqref{eq:leading_asym} into the calculation of
the Jacobian, $J(s)$. We find that to the first two leading orders we can set all $r_s$ equal to each other (since they are very close to each other) 
\begin{equation}
\frac{\partial r_{s+\ell-1}}{\partial r_{s+k}}\sim -\frac{1}{r_{s+k}}\frac{2s+1}
{r_s \dots r_{s+\ell-2}}\simeq - \frac{2s}{r_{s}^\ell } \simeq -(1 +g_{2\ell-2}(2s)^{-1/\ell}) +O((2s)^{-2/\ell}
\end{equation}
The first thing we notice is that to do the analysis we want to take $s\to\infty$ first, obtaining the limit Jacobian
\begin{equation}
    J(\infty) =\begin{pmatrix}  0&1&0&0&\dots\\
0&0&1 & 0&\dots\\
\ddots &\ddots &\ddots &\ddots &\ddots\\
0& 0& 0& \dots &1\\
-1&-1&-1& \dots&-1
    \end{pmatrix}
\end{equation}
This is a $(\ell-1)\times (\ell-1)$ matrix.
Now we do perturbation theory in the corrections proportional to  $1/{2s}^{1\ell}$.
The eigenvalues of $J(\infty)$ are all the different $\ell$-th roots of unity except $1$. The eigenvectors for the eigenvalues $\exp(i c/\ell)$ are given by
\begin{equation}
    \vec w_{c}= \begin{pmatrix}
        1\\
        \exp(i c /\ell)
        \\
        \exp(2 i c /\ell)
        \\
        \vdots
        \\
        \exp((\ell-2) i c /\ell)
    \end{pmatrix}
\end{equation}
and to check this we use the identity
\begin{equation}
\exp((\ell-1) i c /\ell)=-\sum_{k=0}^{\ell-2} \exp( i k c /\ell)
\end{equation}
which can be written as 
\begin{equation}
    x^{\ell-1}+\dots +x+1=0
\end{equation}
where $x$ is a non-trivial $\ell-th$ root of unity. 
At the next order we get that
\begin{equation}
    J(s) \simeq \begin{pmatrix}  0&1&0&0&\dots\\
0&0&1 & 0&\dots\\
\ddots &\ddots &\ddots &\ddots &\ddots\\
0& 0& 0& \dots &1\\
(-1-\epsilon)&(-1-\epsilon)&(-1-\epsilon)& \dots&(-1-\epsilon)
    \end{pmatrix}
\end{equation}
where $\epsilon=g_{2\ell-2}(2s)^{-1/\ell}$. Now we want to estimate the eigenvalues of $J(s)$ for $\epsilon$ small (asymptotically in $s$).

It is convenient to do the first few examples by hand.
For $2\times 2$ matrices we get that the two asymptotic eigenvalues are
\begin{eqnarray}
\lambda_1&=&-\frac{1}{2}+\frac{i \sqrt{3}}{2}-\frac{\epsilon}{2}+\frac{i \epsilon}{2 \sqrt{3}}\\
\lambda_2&=&-\frac{1}{2}-\frac{i \sqrt{3}}{2}-\frac{\epsilon}{2}-\frac{i \epsilon}{2 \sqrt{3}}
\end{eqnarray}
And we will get that the map is expanding, meaning  $|\lambda_{1,2}|>1$, if the following two conditions are satisfied
\begin{eqnarray}
\left|\arg\left[\frac{(-\frac{\epsilon}{2}+\frac{i \epsilon}{2 \sqrt{3}})}{(-\frac{1}{2}+\frac{i \sqrt{3}}{2})}\right]\right |&<& \pi/2\\
\left|\arg\left[\frac{(-\frac{\epsilon}{2}-\frac{i \epsilon}{2 \sqrt{3}})}{(-\frac{1}{2}-\frac{i \sqrt{3}}{2})}\right]\right |&<& \pi/2
\end{eqnarray}
This is equivalent to $|\arg(\epsilon)|=|\arg(g_{4})|<\pi/3$, so that instead of half the complex plane (like for $1\times 1$ matrices), we get convergence on a slice of the complex plane that covers only a third of the complex plane. This slice is symmetric about the real line and requires that $\Re e[g_4]>0$.

A similar calculations with $3\times 3$ matrices show that the eigenvalues are
\begin{eqnarray}
    \lambda_\pi&=& \exp[i \pi]-\frac{\epsilon}{2}\\
     \lambda_{\pi/2}&=& \exp[i\pi/2]+\left(-\frac{1}{4}+\frac{i}{4}\right) \epsilon\\
    \lambda_{-\pi/2}&=& \exp[-i\pi/2]+\left(-\frac{1}{4}-\frac{i}{4}\right) \epsilon
\end{eqnarray}
and it is the two complex ones (the two that are closer to $1$) that determine the bound $|\arg(\epsilon)|<\frac{\pi}{4}$, so that $|\arg(g_6)|<\frac\pi4$ and we only cover $1/4$ of the complex plane, symmetric around the real line with $\Re e[g_6]>0$.

It is easy to convince oneself that the generic case, where the polynomials are of degree $2\ell$, that the system should have the correct expanding property of the eigenvalues when $|\arg(g_{2\ell-2})|<\pi/\ell$, so that the maps are expanding in a simple slice of the complex plane.

The second thing we notice is that since for large $s$ all the matrices approximately commute (meaning that  their eigenvectors are aligned), that we can estimate the total growth by multiplying eigenvalues as if all the matrices have the same eigenvectors. 
We find this way that we get a similar integral estimate for the growth and that all directions expand with a scaling controlled by $\exp(O(g_{2\ell-2}(2s)^{(\ell-1)/\ell}))$, giving us that knowing the asymptotics of $b_s$, we get an error of order 
\begin{equation}
\delta r_{0} \simeq \exp\left[- O(g_{2\ell-2}(2s)^{(\ell-1)/\ell})\right]
\delta r_{s},
\end{equation}
so that errors in the determination of the initial data from the procedure are exponentially suppressed (in the sense that they are smaller than any inverse power laws  of $(2s)^{-f}$). Also $\delta r_s$ is of order of the asymptotic truncation which is in our case $1/(2s)^{1/\ell}$.

In the case of the positive bootstrap, we expect convergence when the leading coupling constant that cannot be removed by scaling (for polynomials of degree $2\ell$ this is $g_{2\ell-2}$) is positive.

The critical case where the leading non-trivial term vanishes is also interesting. In that case we need to expand to higher order and we seem to find that it should also converge if (for example) $g_{2\ell-2}=0$ and $\Re e( g_{2\ell-4})>0$. The slice of the complex plane argument seems to also work in this case, but the convergence is slower, and instead of order 
$exp\left[- O(g_{2\ell-4}(2s)^{(\ell-2)/\ell})\right]$. 

\begin{figure}[ht]
\includegraphics[width=8cm]{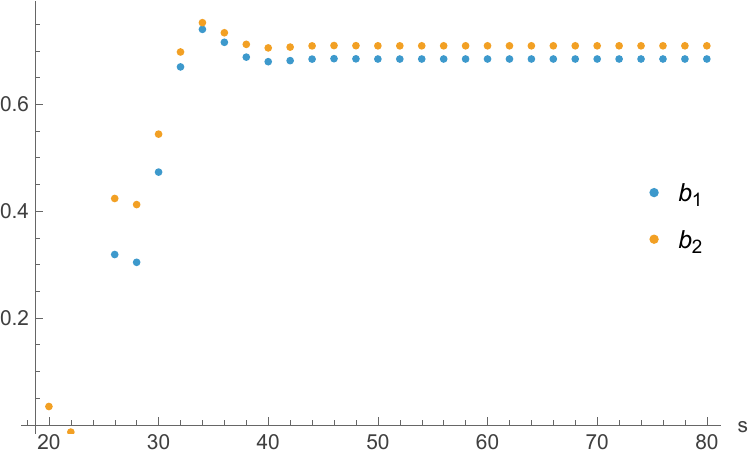}
    \caption{Asymptotic bootstrap method convergence for the case of the sextic potential $x^6/6+5x^4/4-5x^2/2$. The numerical values of $b_1=0.684544569\dots$, $b_2=0.709230991\dots$ evaluated with the NIntegrate subroutine in Mathematica to 100 digits of precission are treated as the {\em correct} values to compare to. }\label{fig:Sextic_convergence}
\end{figure}

\begin{figure}[ht]
\includegraphics[width=8cm]{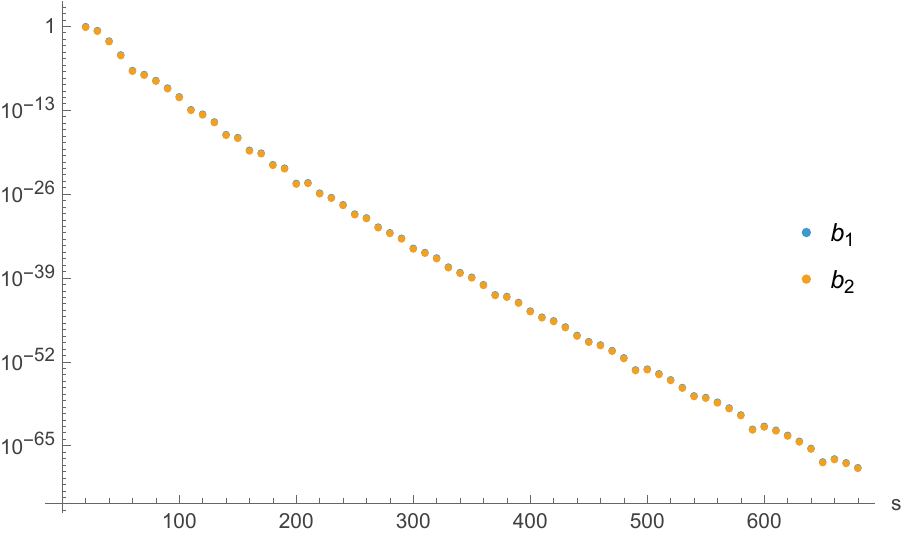}
    \caption{Asymptotic bootstrap method error for the case of the sextic potential $x^6/6+5x^4/4-5x^2/2$. The numerical values of $b_1=0.684544569\dots$, $b_2=0.709230991\dots$ evaluated with the NIntegrate subroutine in Mathematica to 100 digits of precission are treated as the {\em correct} values to compare to. As can be seen, our method at $s\simeq 680$ has more than 65 digits evaluated correctly.
    }\label{fig:Sextic_error}
\end{figure}

We now check on some numerical examples to verify the 
exponentially fast convergence. This is seen for example in figures \ref{fig:Sextic_convergence} and \ref{fig:Sextic_error}. In order to compare the performance of our code relative to the direct numerical integration in Mathematica, we obtain that to evaluate both values together at $s\simeq 800$ takes our code (implemented in Mathematica) approximately $t\sim 0.205861 s$ on a laptop, whereas the numerical integrals $a_{2k}=\int x^{2k}d\mu $ for $k=0,1,2$ to 100 digits of precision to compute the ratios take, evaluated on the same machine using the NIntegrate routing,  $0.095645 s, 0.105499 s$ and $0.113132 s$  respectively. Their ratios $b-2=a_4/a_0, b_1=a_2/a_0$ are the predictions of our numerical method.
In order to get to 300 digits of precision we need to go to $s\simeq 4600$, and for these values it takes
1.14 s to evaluate both $b_{1,2}$ together. The numerical integration subroutines to 400 digits precision takes a similar amount of time.
 
\begin{figure}[ht]
\includegraphics[width=8cm]{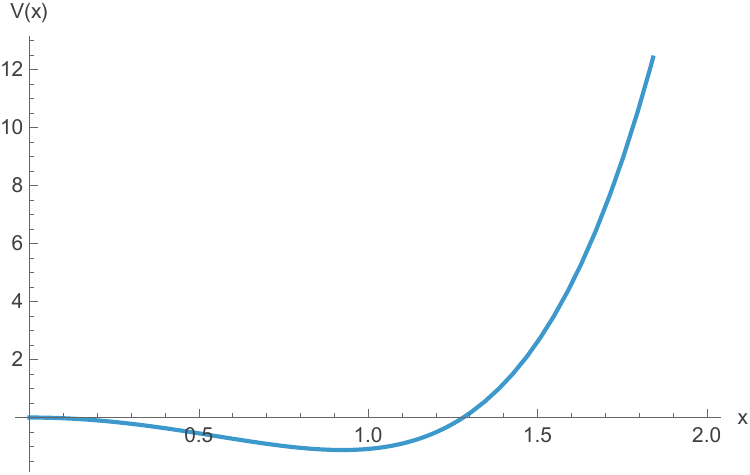}
    \caption{Graph of the potential 
    $V(x)=x^6/6+5x^4/4-5x^2/2$. We chose it because it has a shallow double well (the minimum potential is $V_{min}\sim -1.119$), where certain saddle point methods would produce large corrections from the shallow maximum. 
    }\label{fig:Sextic_potential}
\end{figure}

What this means is that in practice our method can compete with state of the art numerical integration methods for the same problem. The potential $V(x)$ is shown in figure \ref{fig:Sextic_potential}. A similar calculation changing only the sign of $g_2$, for $s=800$ gives the correct answer to over 100 digits of precision. The large difference with the double well answer means we are not sufficiently in the asymptotic regime and the subleading corrections to the estimate of the error are large. For our purposes, the results here already show that the numerical method converges very quickly. The simple timing experiment also shows that the method is very efficient. 
Understanding better this numerical issue is beyond the scope of the present paper. Our goal, after all, was to prove  convergence, which we furthermore found to be exponentially fast. 

\section{Application to Hermitian matrix models}\label{sec:applications}

In this section our goal is to apply our numerical method to the problem of understanding the phase diagram for Hermitian matrix models when one moves from a one cut solution to two cuts for example. Hermitian matrix models are ubiquitous in mathematics and physics. Their most important application is that the statistics of random matrices are the same as that of standard chaotic dynamical systems, exemplified by the distributions and correlations between eigenvalues studied by Wigner and Dyson \cite{wigner1993characteristic,Dyson:1962es,Dyson:1970tza}. This problem of quantum mechanics of chaos is reviewed in \cite{DAlessio:2015qtq}. 
  Random matrices have appeared  in the study of JT gravity \cite{Saad:2019lba}, and in the study of supersymmetric gauge theories \cite{Dijkgraaf:2002vw} just to name a few relatively recent applications to quantum field theory and gravity.
  
For our purposes,  matrix models serve as toy models of statistical mechanics where phase transitions can be more easily understood and that is what we will pursue here as an application of our methods.

In \cite{Lin:2020mme}, Lin proposed that Hermitian  matrix models could be solved with the bootstrap method by
utilizing the loop equations of the matrix models plus positivity. This idea follows the work \cite{Anderson:2016rcw} for unitary matrix models which had the same philosophy.
This method is particularly simple in the case of one-matrix models. One matrix models are usually solved at large $N$ by using a spectral curve \cite{Brezin:1977sv} by introducing a resolvent function. The number of cuts of the spectral curve serves as a way to characterize the different phases of the matrix model.   This positivity method was further studied in \cite{Kazakov:2021lel} where it was argued that the method  would solve such matrix models in all possible situations where the densities of eigenvalues on the cuts (these are all on the real axis) are positive (see also \cite{Kovacik:2025qgj} for additional work in this direction).

We use the method of orthogonal polynomials \cite{Bessis:1979is} and \cite{Itzykson:1979fi} which works for finite $N$ and utilizes the moment
integrals we have been computing at intermediate steps. It is non-perturbative in $N$ (it does not start from the large $N$ expansion of 't Hooft \cite{tHooft:1973alw}).

We start from the standard matrix model problem given by the integral over Hermitian $N\times N$ matrices
\begin{equation}
\int d^{N^2} X \exp( - N \hbox{Tr}(W(X)))\label{eq:mm}
\end{equation}
where we want to take $N\to \infty$ keeping $W(X)$ fixed and we think of the matrix model problem as a (positive) statistical measure. Our goal is to compute the density of eigenvalues $\rho(x)$ associated to the distribution.
As is well known, if $W$ is real, the density $\rho(\lambda)$ has compact support and one should find that the moments of $\rho$
coincide with the moments of the traces of the matrix $X$:
\begin{equation}
    \int_{-\infty}^\infty \lambda^n\rho(\lambda) d\lambda = N^{-1}\langle  \hbox{Tr} X^n\rangle
\end{equation}
where the factor of $N$ in front of the trace is chosen so that the large $N\to\infty$ limit of the right hand side exists. 

The integrals for the ($U(N)$ gauge invariant) traces of $X^n$ with respect to the measure \eqref{eq:mm} can be reduced to statistical mechanics for an eigenvalue measure given by
\begin{equation}
    \int d^n \lambda \prod_{j<k}(\lambda_j-\lambda_k)^2
    \exp\left[ -\sum_k N W(\lambda_k)\right]\label{eq:MM_identity}
\end{equation}
where the product $\prod_{j<k} (\lambda_j-\lambda_k)^2$ is the square of the Vandermonde determinant 
\begin{equation}
 \prod_{j<k} (\lambda_j-\lambda_k)= \det\begin{pmatrix} 1&1&\dots\\
 \lambda_1 &\lambda _2 & \dots\\
 \lambda_1^2 &\lambda_2^2& \dots\\
 \vdots &\vdots &\ddots &\ddots\end{pmatrix},
\end{equation}
and $N$ is the number of eigenvalues. If we think of \eqref{eq:MM_identity} as an integral over a joint  probability density for the eigenvalues (this can be done if the potential is real), then the integral can be normalized to one  and one should study normalized correlators. This can be absorbed in a shift of $W$ by a constant if we want to.

The problem simplifies with the introduction of orthogonal polynomials $P_k(\lambda)$, defined by $P_k(\lambda)=\lambda^k+\dots$, where they are orthogonal with respect to the single eigenvalue potential
\begin{equation}
    \int_{-\infty}^\infty d\lambda \exp(- V(\lambda)) P_k(\lambda)P_{k'}(\lambda) = \tilde h_k \delta_{k, k'}
\end{equation}
where $V(\lambda)=NW(\lambda)$ and $\tilde h_k$ need to be computed.
The matrix model solution and correlators can be written nicely in terms of these polynomials and the factors $h_\ell$. We start with $P_0(\lambda)=1$, 
$P_1(\lambda)= \lambda +c$, etc.
Via row manipulations in the determinant, it is easy to show that
\begin{equation}
\prod_{j<k} (\lambda_j-\lambda_k)=  \det\begin{pmatrix} P_0(\lambda_1) &P_0(\lambda_2)&\dots\\
 P_1(\lambda_1) &P_1(\lambda _2) & \dots\\
 P_2(\lambda_1) & P_2(\lambda_2)& \dots\\
 \vdots &\vdots &\ddots &\ddots\end{pmatrix}
\end{equation}

These polynomials are obtained by first computing the moments of the measure $a_n=\int d\lambda \lambda^n \exp(- N W(\lambda))$.
One then uses these to form a Hankel matrix,  giving exactly the  matrix $M$ as given in \eqref{eq:Hankel_mat}. Passing from the Hankel matrix to the orthogonal polynomials is a numerical problem that suffers from numerical instabilities \cite{gautschi1982generating,gautschi1985orthogonal} (see also \cite{gautschi2004orthogonal}). 
These can all be  traced to the poor conditioning of the moment matrix, which was argued to be the same reason that  makes the numerical bootstrap converge quickly \cite{Berenstein:2025itw}. This is unavoidable. The cure is to have the moments appearing in the Hankel matrix to be evaluated to very high precision (hundreds or thousands of  digits) so that the poor conditioning issue is addressed by having enough precision and the "floating point" truncation errors are under control. For completeness, we add  basic information on orthogonal polynomials in appendix \ref{sec:orthogonal}.

Positivity of $M$ implies that it can be interpreted as an inner product. One then performs a Gram-Schmidt algorithm to determine the orthogonal polynomials.
The positive bootstrap method uses positivity of this inner product (the Hankel matrix) to find bounds.
These polynomials also satisfy a recurrence equation, which we  use to compute them more readily from the Hankel matrix.
To find the recursion, one needs to compute the $h_k$ and invert them (essentially, in this step on is performing an inverse of  the matrix $M$). Here is where the numerical instability is appearing: the ratio of the largest eigenvalue to the smallest eigenvalue is exponentially large in the size of the matrix. This is the main reason that we need to determine the matrix $M$ to very high precision on the entries.

To make contact with the previous section, we are interested in the case that $W(\lambda)$ is an even polynomial such that our method that was specialized to even potentials applies. 

Let $NW(\lambda)= V(x)$, where we choose
$W(\lambda) = \sum_{n=1}^\ell w_{2n} \lambda ^{2n}/{2n}$ to be even and we also use a change of variables to $x= N^{1/(2\ell)}\lambda $.
This corresponds to a potential $V(x)$ given by
\begin{equation}
V(x)= \sum_{n=1}^\ell \frac{g_{2n}}{2 n} x^{2n}
\end{equation}
with $g_{2n}= N (N^{-n/\ell}) w_{2n}$, or equivalently $w_{2n}= g_{2n} N^{n/\ell -1} $.

We want to work in the simplest setting where the matrix model is not solved analytically by standard special functions (Bessel functions). This requires a sextic potential.

If we want to work with a sextic potential and we fix $w_2,w_4$, after a rescaling we will get $g_{2,4}$. Convergence of our method requires (when the potential is  real) that $g_4>0$, whereas $g_2$ can have either sign. When $g_2$ is very large and  negative, we expect a two cut solution, whereas if $g_2$ is very large and positive we expect a one cut solution. We will look for the transition point in $g_2$ from one cut to two cuts, again to illustrate our technique.
We will also fix $N=100$ for simplicity, so we do not verify the large N scaling by varying $N$ directly. We just assume it. We can also double check this by solving for the spectral curve of the corresponding problem and comparing both to each other.
The importance of the $N$ rescaling has to do with how we evaluate the matrix model solution at large $N$ using other methods, not with the orthogonal polynomials. 
In what follows, we use the $x$ variables and fix $g_4=5$.
Similarly, we compute the orthogonal polynomials associated to $V$ for the $x$ variables.

Given the polynomials $P_k(x)$ (these are the same as the $P_k(\lambda)$ up to the rescaling of the variables), the density of eigenvalues is
\begin{equation}
\rho(x) ={\cal N}\frac 1N\sum_{k=0}^{N-1} \frac{P_k(x)^2}{ h_k} \exp(-V(x))\label{eq:density}
\end{equation}
where the $h_k$ are similar to $\tilde h_k$, but adapted to the measure of $x$, rather than the large $N$ measure of $\lambda$. We get
\begin{equation}
  {\cal N}  \int_{-\infty}^\infty dx \exp(- V(x)) P_k(x)P_{k'}(x) =  h_k \delta_{k, k'}
\end{equation} 
and ${\cal N}$ is the normalization constant defined in \eqref{eq:moments} that makes it so that $b_0=1$. 
It is easy to then prove that the $\int_{-\infty}^\infty \rho(x) dx=1$ by direct substitution.

\begin{figure}[ht]
\includegraphics[width= 5 cm]{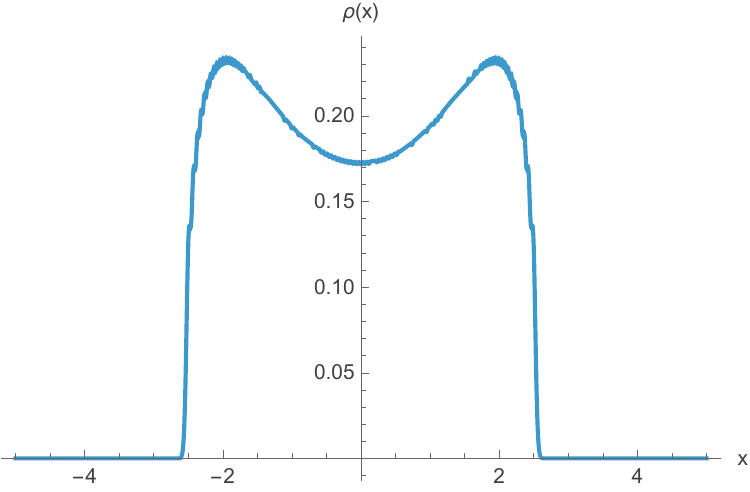}
\includegraphics[width= 5 cm]{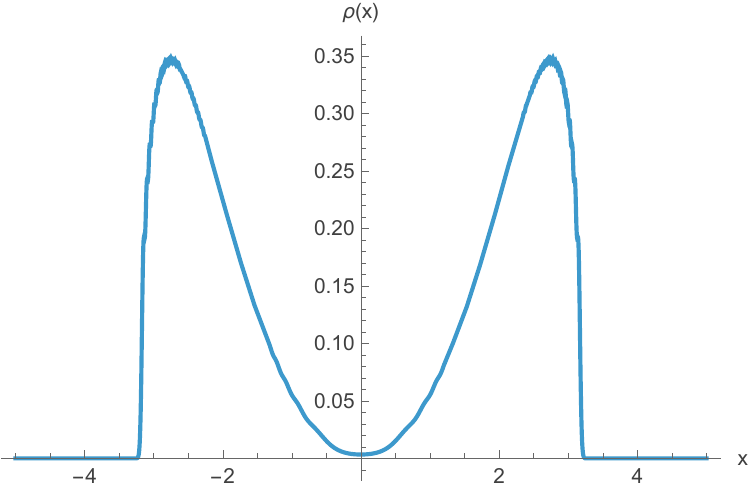}
\includegraphics[width=5cm]{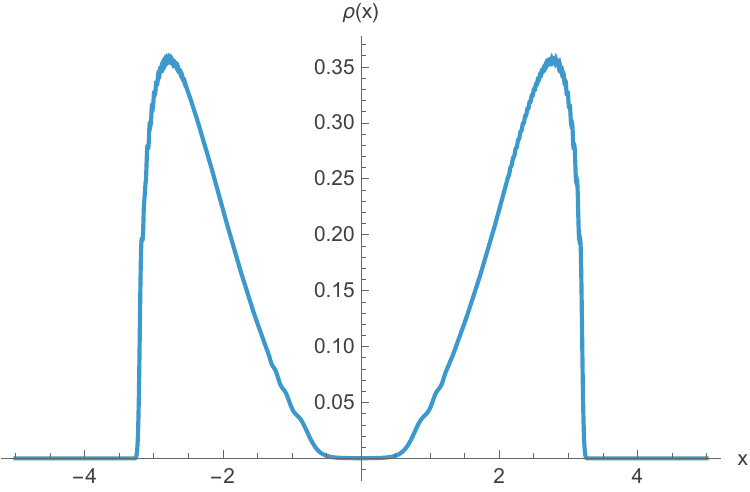}
\caption{Eigenvalue densities for the sextic potential 
$V(x)= \frac{g_2}{2} x^2+ \frac 5 4 x^4 +\frac 16 x^6$ with $g_2=10, -64.-68$ respectively at $N=100$. The transition point from one to two cuts is close to $g_2=-64$.On the left we see a clear one cut solution and on the right we see a clear two cut solution (there is a region of tiny eigenvalue density support in the middle). }\label{fig:eival_density}
\end{figure}

We don't have a direct way to calculate ${\cal N}$ with what we have described in this paper (this is something we are looking into), so that when it is  needed to evaluate $\rho$, the numerical constant is evaluated by numerical integration in Mathematica to as high precision as is necessary. Incidentally, for the evaluation of the Polynomials $P_k$, in some calculations we have found that to numerically evaluate the orthogonal polynomials in \eqref{eq:density} we use up to 10000 digits of precision in intermediate steps, and we chose to evaluate  ${\cal N}$ and the polynomials themselves to about 900 digits of precision. In the polynomials there are large cancellations that can only be dealt with with high precision values for the coefficients. Our results of this procedure are shown in \ref{fig:eival_density}.

We clearly see a transition from one cut to two cuts. To produce the last plot at $g_2=-68$ we solved the system starting with a value of $s=(32^3)/2$ which was chosen so that the cube root appearing in $\chi_{s}\simeq (2s)^{1/3}-g_{4}/{3}$ is an integer. We used  10000 digits of precision in intermediate steps \footnote{It should be noted that since $\chi_s$ as written only depends on $g_4$, we expect that a better asymptotic determination of the $r_s$ will receive higher order corrections that depend also on $g_2/(2s)^{1/3}$ and since $g_2$ is large these can be substantial. It is clear that improving $\chi_s$ will lead to faster convergence for lower values of $s$, so that taking $s$ this large is not necessary. Such large value of $s$ requires in this case multiplying $3\times 3 $ matrices $s$ times.}. It should also be noticed that
even though there is a similar transition in the case of the quartic potential model, it also requires solving the model with $g_2<0$, which is in the region of our method where the method does not converge (this can always be fixed by adding a small sextic piece and taking it to zero eventually, but that is not a direct solution).

\begin{figure}[ht]
\includegraphics[width=8cm]{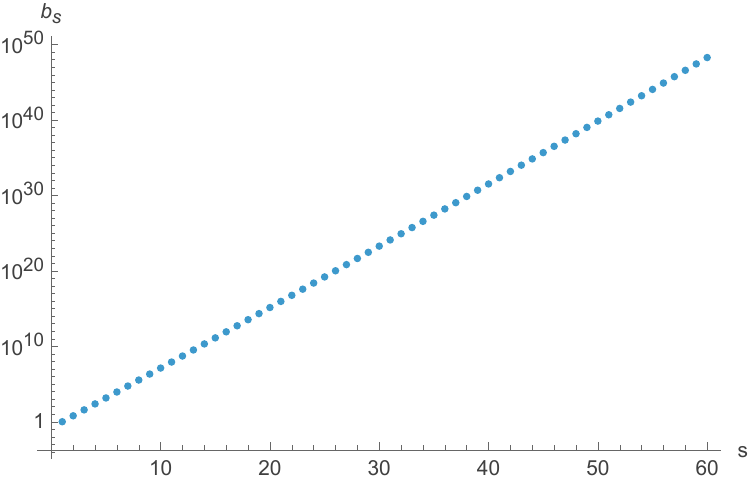}
\caption{First few moments of the problem with potential 
$V(x)= -34 x^2+ \frac 5 4 x^4 +\frac 16 x^6$}\label{fig:moments}
\end{figure}

For example, the first few moments of the corresponding 
distribution are shown in figure \ref{fig:moments}. At the same time, the smallest eigenvalue of the $100\times 100$ principal minor of the corresponding Hankel matrix is roughly $\lambda_{min}\simeq 1.38\times 10^{-59}$, giving us a poorly conditioned matrix with condition number of  order $\kappa\simeq 10^{50+59}$. It only gets worse for larger matrices. It is exactly this poor conditioning that requires large precision in all computations. Here we have made no effort to improve the numerics by rescaling $x$ to reduce the conditioning problem of the Hankel matrix $M$. 

We can trace some of the growth of the moments to the fact that the minimum of the potential for $g_2=-68$ is located at $x_0\simeq 2.47$, 
so the even moments must grow at least as fast as $b_s \simeq x_0^{2s}$. Since we have 100 orthogonal polynomials to consider, we need to be able to handle cancellations up to the level of the conditioning of the Hankel matrix $M$ with comfort. This requires at least 200 digits of precision in the entries of the matrix $M$ and we make certain to use more. We also found that since we needed to go to very large $s$ for the computations to converge for $b_2,b_4$, that in this case we needed to be extra careful with the precision of the numerical method.

For illustration purposes, here is the 29-th orthogonal polynomial
\begin{eqnarray}
P_{29}(x)&=& x^{29}-81.07085388 x^{27}+3040.493271 x^{25}-69916.23905 x^{23}+1.101212345\times 10^6
   x^{21}\nonumber\\&-&1.256681575\times 10^7 x^{19}
   +1.071473478\times 10^8 x^{17}-6.933542557\times
   10^8 x^{15}\nonumber \\&+&3.421663872\times 10^9 x^{13}\nonumber-1.281406989\times 10^{10}
   x^{11}+3.584584002\times 10^{10} x^9\\&-&7.262753337\times 10^{10} x^7+1.007446636\times
   10^{11} x^5\nonumber-8.563394509\times 10^{10} x^3\\&+&3.364973460\times 10^{10} x
\end{eqnarray}
which shows that there are very large coefficients in the polynomials $P$ and there must be cancellations to 
high order just evaluating $P$ in the region of interest. Also, for example, the maximum coefficient for $P_{99}(x)$ is $1.3953\times 10^{37}$ showing again how the numerical issues are very delicate at all the steps in the process of finding the 
solution via the orthogonal polynomials.

Now let us check that our numerical method agrees with the large $N$ solution of the matrix model in the leading large $N$ approximation. In order to do that, we need to pass from $V(x)$ to $W(\lambda)$ and express the spectral curve that solves the matrix model in terms of $W(\lambda)$. The coefficients for $W$ are then
$w_2= g_2/N^{2/3}$ and $w_4=5/N^{1/3}$ (after all, we are fixing $g_4=5$ in our plots).

The spectral curve  is given by the resolvent
\begin{equation}
    R( \lambda)= \frac 1 N \hbox{Tr} \frac{1}{\lambda-X}
\end{equation}
and it satisfies the one cut ansatz equation
\begin{equation}
    R= -\frac{W'}{2}+ G(\lambda) \sqrt{\lambda^2-a^2}
\end{equation}
where $G$ is a polynomial of degree four and $\pm a$ are the branching points of the curve. In the appendix \ref{sec:spec} we give more details.
Since $R(\lambda)\to 0$ as $\lambda\to \infty$, we can expand for large $\lambda$ and compare the coefficients of the $G$ multiplied by $\sqrt{\lambda^2-a^2}$ to $W'/2$. We remove all the terms in $R$ that grow. There is also the normalization condition $R(\lambda)\to 1/\lambda$ that arises from $\hbox{Tr(1)}=N$. These are enough to solve for the coefficients of $G$ and produce an algebraic equation of degree $6$ that $a$ needs to solve.
This lets us reduce the problem of finding $a$,  to solving a cubic equation for $a^2$ and giving us a determination of the position of the cut. Once we have the location of the cut in the $\lambda$
coordinate, we pass to $\tilde a= N^{1/6}a$ for the location of the cuts in the $x$ variable.

\begin{figure}
    \includegraphics[width=6 cm]{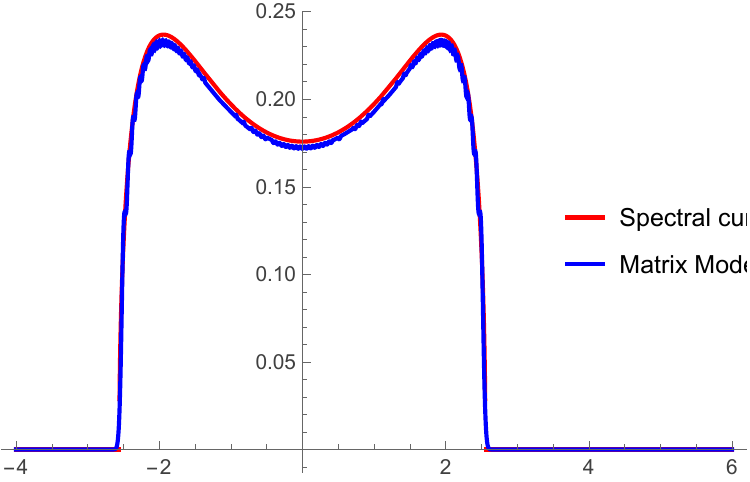}
    \includegraphics[width=6 cm]{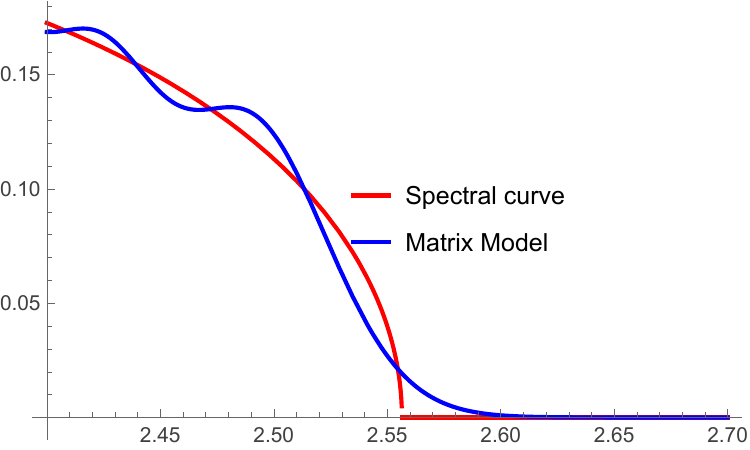}
    \caption{Comparison of the eigenvalue density between  the spectral curve computation at $g_2=10,g_4=5$ and our method using orthogonal polynomials. The scaling of the spectral curve result has been exaggerated by $2\%$ to show that both curves are really close to each other. Without the rescaling they are hard to distinguish by visual inspection. On the right we zoom in to the edge of the cut, making it possible to see the finite $N$ deviations from the square root cut.}\label{fig:cut-comp}
\end{figure}

\begin{figure}
    \includegraphics[width=8 cm]{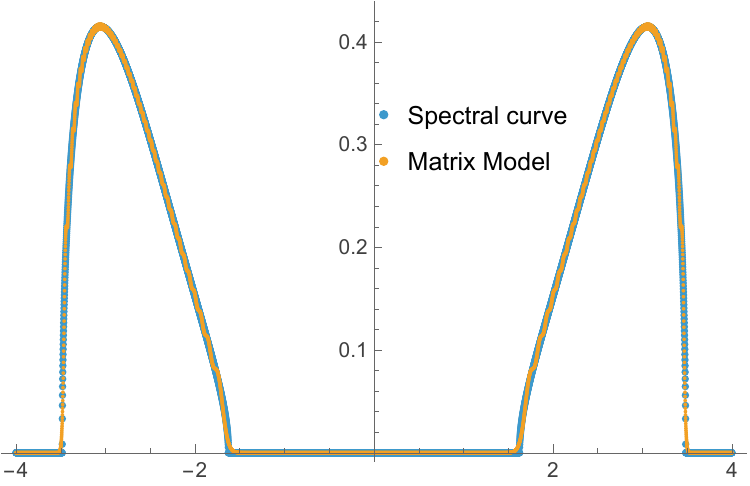}
    \caption{Comparison of the eigenvalue density between  the spectral curve computation at $g_2=-88 ,g_4=3$ and our method using orthogonal polynomials. We clearly see that the two cut solution and the density computed from the polynomials agree. 
    }\label{fig:2cut-comp}
\end{figure}

For $g_2=10$, we find that $\tilde a\simeq 2.556$ clearly consistent with \ref{fig:eival_density}. Similarly, for $g_2=-64$, we find the position of the cut at $\tilde a\simeq 3.193$, also consistent with our plots. We can also compare the density of eigenvalues computed from the discontinuity  of the cut of the spectral curve to the matrix model (we have rescaled the spectral curve to the correct $x$). This is shown in figure \ref{fig:cut-comp}. 
It works also for the marginal case $g_2=-64$, where the density of eigenvalues computed from the one cut spectral curve is still slightly positive at $x=0$.
As for the two cut case, we show an example in \ref{fig:2cut-comp} where again, the spectral curve method and the finite $N$ method based on orthogonal polynomials coincide.

We find a very close match of the computation with orthogonal polynomials and the equivalent computation with the discontinuity of the cut of the spectral curve. We have  slightly displaced them relative to each other to see that both curves have the same form.
We have also zoomed in to the region where the cut ends. This is an interesting region as the eigenvalues can penetrate (tunnel) into the forbidden region. This is a non-perturbative process that requires  $N$ finite o more precisely, double scaled to the edge of the distribution (for a recent discussion, see \cite{Johnson:2022wsr} and references therein). 

\section{Conclusion}\label{sec:concl}

In this paper we have presented an asymptotic
bootstrap method to evaluate moments of a measure on the real line. These measures are of the type
\begin{equation}
    \int d\mu \simeq \int_{-\infty}^\infty dx \exp(-V(x))\label{eq:problem2}
\end{equation}
with $V(x)$ a polynomial potential with real coefficients and whose leading coefficient is chosen so that the total integral is finite. In this paper, for simplicity we used an even potential $V(x)=V(-x)$, but we do not believe that this is a necessary condition for the method to work.

The method relies on two ingredients: a recursion relation between moments of the measure, and we also use some asymptotic information about the ratios of successive moments
(this is a generalization of a simpler condition that appeared in circle measure problems \cite{Berenstein:2025itw}).
With this information the asymptotic behavior can be back-propagated to the moments that initialize the recursion relation by solving a linear algebra problem with high precision arithmetic.

We used this method to give a proof of convergence of the positive bootstrap method for the same problem in a restricted setting. The idea is that some of the positivity inequalities  are enough to get an approximation to the asymptotic behavior, so that when the asymptotic method converges, so does the positive method.

The asymptotic method also works in cases where the positive method doesn't: when we extend the coefficients of $V(x)$ to the complex plane. These would be cases where the integral has a sign problem (it can be highly oscillating).

Numerically, when it converges, the asymptotic method is  very efficient, especially compared to the positive bootstrap method. The positive method uses a semi-definite programming algorithm which utilizes a lot of computational resources when one needs to move to high precision and large matrices. 
Our numerical method also seems to be able to compete in execution time with high precision numerical integration methods that would compute the same moments of the distribution. Understanding better why this method seems to be very efficient is an interesting avenue to explore.  

Our method evaluates the ratio of the moments of the measure very efficiently, but we have not determined the total normalization
of the integral in equation \eqref{eq:problem2}. It would be interesting to see if this is possible.

The region of convergence is specified as follows
\begin{equation}
V(x)= \frac {x^{2\ell}}{2\ell}+g_{2\ell-2}\frac {x^{2\ell-2}}{2\ell-2}+\dots
\end{equation}
and we require that $|\arg(g_{2\ell-2})|< \pi/{\ell}$.
In particular, this implies that $\Re e(g_{2\ell-2})>0$. In the case where we can restrict to the positive bootstrap case, the method converges on half of the allowed parameter space. 
This is the region where the leading non-trivial coefficient is positive.

We applied this method to the problem of understanding the phase structure of matrix models of a single matrix. We were able to show that we could see transitions from one-cut to two cut solutions for real potentials (in the example we studied a  sextic potential and the quadratic term could take either sign). The need for high precision arises from the method we used: we computed the orthogonal polynomials of the measure numerically. This problem is well known and it requires high precision since computing the orthogonal polynomials can have numerical instabilities related to a large hierarchy of eigenvalues in a matrix built from the moments of the distribution. This matrix is effectively inverted in the process of building the orthogonal polynomials. 

Similarly, we can consider the same problem of evaluating moments of the expression
\begin{equation}
\int_\infty^\infty dz \exp(-z^2/2-g z^4/4)
\end{equation}
for $g$ small so long as it is positive. In that case we know that we can expand in Feynman diagrams and we get an asymptotic series (see for example the lectures \cite{Marino:2012zq}). Our numerical method works very well even in the case of g small (we need a change of variables $x=g^1/4 z$, so the coupling is moved to the term $\exp(-x^2/2 g^{1/2})= \exp(-g_2 \frac {x^2}2)$ with $g_2= g^{-1/2}$). The error we get is of order $\exp(-g_2 \sqrt{2s})$
which is non-perturbative and depends on a truncation parameter of the method (the asymptotic moment  $s$   where we compare to the asymptotic behavior, which is smaller than any power of $g$). Our method works even when $g_2$ is complex and with positive real part (in the regime we want it is very large).
In that sense we would be able to perform highly efficient numerical evaluations of functions with asymptotic series where resurgence and instanton methods are used to do computations. For recent work on matrix model applications of these transseries methods see \cite{Kager:2026cod} and references therein. Our methods at finite $N$ would do some of that work.

We can hope that since our method depends on asymptotics, that it might have some similarities to these asymptotic series problems and that our restrictions on convergence might have some relation to phenomena that appear in that other setting like the presence of anti-Stokes lines. These lines are used to determine when non-perturbative effects become important. 
It seems that the recursive method we have also works for integrals in the complex plane that are not tethered to the real axis, so long as they are in some steepest descent contour that renders them convergent. We leave the exploration of this idea to future work.

On another note,  in matrix models where there are multi-cut solutions to the loop equations, these usually depend on some filling fractions. The orthogonal polynomials derived from the moment problem associated to the measure \eqref{eq:problem} provide only one such filling fraction solution. We would like to be able to produce the other ones somehow (this is usually handled with other methods \cite{Kovacik:2025qgj} that also use positivity in some steps).

We would also like to point out that this problem arose from trying to understand the quantum mechanical bootstrap problem formulated by \cite{Han:2020bkb} et al. It would be useful to see if our proof can be extended to efficiently computing the spectrum of Hamiltonians in one dimension.

\acknowledgments

We are very grateful to J. Rodrigues and V. Rodriguez for discussions. We also thank N. Levine for feedback on the draft.  The work of D.B. was
supported in part by the Department of Energy under grant DE-SC 0011702.

\appendix

\section{Orthogonal Polynomials}\label{sec:orthogonal}
In this section we will give a brief overview of the concept of orthogonal polynomials.(For a more detailed explanation, we are following the book \cite{gautschi2004orthogonal}). 
\newline 
Let $\mathbb{P}$ be the space of real polynomials. For any pair $u,v$ in $\mathbb{P}$, we can define the inner product for a positive measure $d\lambda$, that has finite moments of all orders, 
\begin{equation*}
    \langle u,v \rangle =\int_\mathbb{R} u(t)v(t)d\lambda(t)
\end{equation*}
If $\langle u,v\rangle =0 $, then $u$ is said to be orthogonal to $v$.   

It can be shown that a set of monic orthogonal polynomials $P_0,...,P_n$ forms a basis of the space of polynomials with degree $\leq n$, $\mathbb{P}_n$. Let $\sum_{k=0}^n\gamma_kP_k\equiv0$, by taking the inner product of both sides with $P_j,j=0,1...,n$ $\gamma_j\langle P_j, P_j\rangle=0$, hence $\gamma_j=0,j=0,1,...,n$ and $P_0,...,P_n$ are linearly independent. Now we will show that any polynomial $p$ of degree $\leq n $ can be written as $p=c_0P_0+c_1P_1+...+c_nP_n$. Taking the inner product on both sides with $P_k$ gives $c_k=\frac{\langle p,P_k\rangle}{\langle P_k,P_k\rangle}$, $k=0,1,...,n$.  Hence, $p-\sum _{k=0}^nc_kP_k$ is orthogonal to $P_0,...,P_n$,therefore, writing $p-\sum _{k=0}^nc_kP_k= a_0+a_1t+...+a_nt^n$    
\begin{equation*}
    0=\langle p-\sum _{k=0}^nc_kP_k,P_n\rangle =a_n\langle t^n,P_n\rangle =a_n\langle P_n,P_n\rangle 
\end{equation*}
Since $\langle P_n,P_n\rangle>0$, we have that $a_n=0$. Similarly, we can show that $a_{n-1}=a_{n-2}=...=0$. 
\newline\newline One of the most important properties of the monic orthogonal polynomials is that they satisfy the three-term recurrence relation: 
\begin{equation}
\label{recurrence}
   P_{k+1}(t)=(t-\alpha_k)P_k(t)-\beta_kP_{k-1}(t),\ \ k=0,1,2...,
    \end{equation}
    \begin{equation*}
    P_{-1}(t)=0,\ \ P_0(t)=1
\end{equation*}
where
\begin{equation}\label{alpha_k}
    \alpha_k=\frac{\langle tP_k,P_k\rangle_{d\lambda}}{\langle P_k,P_k\rangle_{d\lambda}},\ k=0,1,2,...
\end{equation}
\begin{equation}\label{beta_k}
\beta_k=\frac{\langle P_k,P_k\rangle_{d\lambda}}{\langle P_{k-1},P_{k-1}\rangle_{d\lambda}},\ k=1,2,...
\end{equation}
Since $P_0,...,P_n$ form a basis of $\mathbb{P}_n$ and $P_{k+1}-tP_k$ is a polynomial of degree $\leq k$, it can be written as: 
\begin{equation}
\label{proof}
P_{k+1}(t)-tP_k(t)=-\alpha_kP_k(t)-\beta_kP_{k-1}(t)+\sum_{j=0}^{k-2}\gamma_{kj}P_j(t)
\end{equation}
for certain constants $\alpha_k,\beta_k$ and $\gamma_{kj}$ where $P_{-1}(t)=0$ and the sums for $k=0$ and $k=1$ are understood to be 0. Taking the inner product of both sides with $P_k$
\begin{equation*}
    -\langle tP_k,P_k\rangle =-\alpha_k\langle P_k,P_k\rangle  
\end{equation*} therefore,
\begin{equation*}
    \alpha_k=\frac{\langle tP_k,P_k\rangle }{\langle P_k,P_k\rangle }
\end{equation*} which proves (\ref{alpha_k}). 
Now we take the inner product with $P_{k-1}(k\geq1)$
\begin{equation*}
    -\langle tP_k,P_{k-1}\rangle =-\beta_k\langle P_{k-1},P_{k-1}\rangle  
\end{equation*}
By the definition of the inner product it is easy to see that $\langle tP_k,P_{k-1}\rangle = \langle P_k,tP_{k-1}\rangle=\langle P_k,P_k+p\rangle$ where $p$ is some polynomial of degree $<k$, hence $\langle tP_k,P_{k-1}\rangle=\langle P_k,P_k\rangle$, and thus: 
\begin{equation*}
    \beta_k=\frac{\langle P_k,P_k\rangle}{\langle P_{k-1},P_{k-1}\rangle}
\end{equation*} and this proves (\ref{beta_k}). 
Lastly we can prove (\ref{recurrence}) by taking the inner product with $P_i,i<k-1$
\begin{equation*}
    -\langle tP_k,P_i\rangle =\gamma_{ki}\langle P_i,P_i\rangle 
\end{equation*}
We use again that $\langle tP_k,P_i\rangle=\langle P_k,tP_i\rangle$ and since is $tP_i$ of degree $<k$, then $\langle P_k,tP_i\rangle=0$ Therefore, $\gamma_{ki}=0$ for $i<k-1$ and we obtain (\ref{recurrence}). 

\section{The spectral curve}\label{sec:spec}
In this section, we will explain how to compute the spectral curve for the large N solution of the matrix model with a sextic potential, $W(x)=\frac{w_2}{2}x^2+\frac{w_4}{4}x^4+\frac{1}{6}x^6$.
\begin{equation*}
    \int d^{N^2}X\exp(-NTr(W(X)))
\end{equation*}

In the large N-limit, we can assume that the eigenvalue density is continuous and initially, we will assume it has compact support symmetric about the origin $[-a, a]$; this is called the one-cut solution. 

In order to easily solve for the spectral curve, we introduce the resolvent, which is a complex function defined as follows (we follow the easily readable review \cite{Anninos:2020ccj}): 
\begin{equation*}
    R_N(z)=\frac{1}{N}Tr\frac{1}{z\mathbb{I}_N-M}
\end{equation*}
and as
 $N$ tends to infinity it becomes: 
\begin{equation*}
    R(z)=\int_{-a}^ad\mu \frac{\rho(\mu)}{z-\mu}
\end{equation*}
where we have introduced an eigenvalue density $\rho(\mu)$ to count eigenvalues.

The resolvent must verify a series of properties \cite{Brezin:1977sv}: 
\begin{enumerate}
\item It is analytic for all $z$ in the interval $(-a, a)$
\item It must decay as $\frac{1}{z}$ as $|z|$ tends to infinity. 
\item  It is real for $z$ outside the interval $(-a, a)$
\item  When $z$ approaches the interval $(-a,a)$ it satisfies: 
\begin{equation*}
    R(x\pm i\varepsilon)=\frac{1}{2}W'(x)\mp i\pi\rho(x)
\end{equation*}
\end{enumerate}
Hence, we can assume the resolvent is of the form: 
\begin{equation*}
    R(z)=\frac{1}{2}W'(z)-P(z)\sqrt{z^2-a^2}
\end{equation*}
where $P(z)$ is some polynomial.

The resolvent also satisfies a quadratic condition that is derived from the loop equations, which are written as follows
\begin{equation}
0=\int d^{N^2 X} \frac{1}{N}\hbox{Tr}\left(
\partial_X \frac 1{z-X}\right) \exp(-N \hbox{Tr}(W(X)))
\end{equation}
This produces a term
\begin{equation}
  \langle  \frac{1}{N}\hbox{Tr}\left(
 \frac 1{z-X} N W'(X)\right)\rangle
\end{equation}
and it is convenient to write $W'(X)= W'(X)-W'(x)+W'(z)$ to show that this term is a polynomial of $z$ of degree $\deg(W)-2$ plus
$W'(z) R(z)$, which includes the resolvent. The other term where $\partial_X$ acts on the resolvent itself gives the resolvent squared. One then uses factorization at large $N$ to find that $R$ satisfies an equation of the form $R(z)^2-W'(z) R(z) + f(z)=0$, which justifies the ansatz (incidentally, this factorization property was essential to match this same problem of matrix models to supersymmetric gauge theories in \cite{Cachazo:2002ry}).

According to property (ii), we need $P(z)$ to be a fourth-degree polynomial since $P(z)\sqrt{z^2-a^2}$ needs to cancel all of the terms in $\frac{1}{2}V'(z)$ that grow polynomially. As a result: 
\begin{equation*}
   P(z)=\frac{1}{2}z^4+\frac{1}{4}(a^2+2w_4)z^2+\frac{1}{16}(3a^4+8w_2+4a^2w_4) 
\end{equation*}
Additionally, the normalization condition provides an equation for $a$: 
\begin{equation*}
    \frac{5a^6}{32}+\frac{a^2w_2}{4}+\frac{3a^4w_4}{16}-1=0
\end{equation*}
Solving for $a$, we obtain the support of the eigenvalue spectral density.
\newline Now we can derive the spectral curve using the following expression \cite{Marino:2004eq}: 
\begin{equation*}
y = W'(z) - 2R(z) 
\end{equation*}
Then, the spectral curve is given by: 
\begin{equation*}
   y^2= \frac{1}{64}(z^2-a^2)(3a^4+4a^2(w_4+z^2)+8(w_2+w_4z^2+z^4))^2
\end{equation*}
The variable $y$ is (up to normalization) just a shifted version of the resolvent, so it stands on the same footing as it. 

We can follow an analogous procedure to obtain the two-cut solution. In this case, we will assume that the spectral eigenvalue density has support $[-b,-a]\cup[a,b]$. Following the same ansatz, the resolvent will be of the form: 
\begin{equation*}
R(z)=\frac{1}{2}W'(z)-P(z)\sqrt{z^2-a^2}\sqrt{z^2-b^2}
\end{equation*}
However, in this case, P(z) is a third-degree polynomial. Similarly, we obtain: 
\begin{equation*}
    P(z)=\frac{1}{2}z^3+\left(\frac{1}{2}w_4+\frac{1}{4}(a^2+b^2)\right)z
\end{equation*}
Besides, we have a system of equations for $a$ and $b$ that can be solved to obtain the support of the eigenvalue spectral density. 
\begin{align*}
    &\frac{a^6}{16}-\frac{a^4b^2}{16}-\frac{a^2b^4}{16}+\frac{b^6}{16}+\frac{a^4w_4}{16}-\frac{1}{8}a^2b^2w_4+\frac{b^4w_4}{16}-1=0
    \\&w_2=\frac{1}{8}(-3a^4-3b^4-4b^2w_4-2a^2(b^2+2w_4))
\end{align*}
Lastly, the spectral curve is:
\begin{equation*}
    y^2=\frac{1}{4}z^2(z^2-a^2)(z^2-b^2)(a^2+b^2+2(w_4+z^2))^2
\end{equation*}

\bibliography{refs}

\begin{thebibliography}{45}%
\makeatletter
\providecommand \@ifxundefined [1]{%
 \@ifx{#1\undefined}
}%
\providecommand \@ifnum [1]{%
 \ifnum #1\expandafter \@firstoftwo
 \else \expandafter \@secondoftwo
 \fi
}%
\providecommand \@ifx [1]{%
 \ifx #1\expandafter \@firstoftwo
 \else \expandafter \@secondoftwo
 \fi
}%
\providecommand \natexlab [1]{#1}%
\providecommand \enquote  [1]{``#1''}%
\providecommand \bibnamefont  [1]{#1}%
\providecommand \bibfnamefont [1]{#1}%
\providecommand \citenamefont [1]{#1}%
\providecommand \href@noop [0]{\@secondoftwo}%
\providecommand \href [0]{\begingroup \@sanitize@url \@href}%
\providecommand \@href[1]{\@@startlink{#1}\@@href}%
\providecommand \@@href[1]{\endgroup#1\@@endlink}%
\providecommand \@sanitize@url [0]{\catcode `\\12\catcode `\$12\catcode
  `\&12\catcode `\#12\catcode `\^12\catcode `\_12\catcode `\%12\relax}%
\providecommand \@@startlink[1]{}%
\providecommand \@@endlink[0]{}%
\providecommand \url  [0]{\begingroup\@sanitize@url \@url }%
\providecommand \@url [1]{\endgroup\@href {#1}{\urlprefix }}%
\providecommand \urlprefix  [0]{URL }%
\providecommand \Eprint [0]{\href }%
\providecommand \doibase [0]{https://doi.org/}%
\providecommand \selectlanguage [0]{\@gobble}%
\providecommand \bibinfo  [0]{\@secondoftwo}%
\providecommand \bibfield  [0]{\@secondoftwo}%
\providecommand \translation [1]{[#1]}%
\providecommand \BibitemOpen [0]{}%
\providecommand \bibitemStop [0]{}%
\providecommand \bibitemNoStop [0]{.\EOS\space}%
\providecommand \EOS [0]{\spacefactor3000\relax}%
\providecommand \BibitemShut  [1]{\csname bibitem#1\endcsname}%
\let\auto@bib@innerbib\@empty
\bibitem [{\citenamefont {Belavin}\ \emph {et~al.}(1984)\citenamefont
  {Belavin}, \citenamefont {Polyakov},\ and\ \citenamefont
  {Zamolodchikov}}]{Belavin:1984vu}%
  \BibitemOpen
  \bibfield  {author} {\bibinfo {author} {\bibfnamefont {A.~A.}\ \bibnamefont
  {Belavin}}, \bibinfo {author} {\bibfnamefont {A.~M.}\ \bibnamefont
  {Polyakov}},\ and\ \bibinfo {author} {\bibfnamefont {A.~B.}\ \bibnamefont
  {Zamolodchikov}},\ }\bibfield  {title} {\bibinfo {title} {{Infinite Conformal
  Symmetry in Two-Dimensional Quantum Field Theory}},\ }\href
  {https://doi.org/10.1016/0550-3213(84)90052-X} {\bibfield  {journal}
  {\bibinfo  {journal} {Nucl. Phys. B}\ }\textbf {\bibinfo {volume} {241}},\
  \bibinfo {pages} {333} (\bibinfo {year} {1984})}\BibitemShut {NoStop}%
\bibitem [{\citenamefont {El-Showk}\ \emph {et~al.}(2012)\citenamefont
  {El-Showk}, \citenamefont {Paulos}, \citenamefont {Poland}, \citenamefont
  {Rychkov}, \citenamefont {Simmons-Duffin},\ and\ \citenamefont
  {Vichi}}]{El-Showk:2012cjh}%
  \BibitemOpen
  \bibfield  {author} {\bibinfo {author} {\bibfnamefont {S.}~\bibnamefont
  {El-Showk}}, \bibinfo {author} {\bibfnamefont {M.~F.}\ \bibnamefont
  {Paulos}}, \bibinfo {author} {\bibfnamefont {D.}~\bibnamefont {Poland}},
  \bibinfo {author} {\bibfnamefont {S.}~\bibnamefont {Rychkov}}, \bibinfo
  {author} {\bibfnamefont {D.}~\bibnamefont {Simmons-Duffin}},\ and\ \bibinfo
  {author} {\bibfnamefont {A.}~\bibnamefont {Vichi}},\ }\bibfield  {title}
  {\bibinfo {title} {{Solving the 3D Ising Model with the Conformal
  Bootstrap}},\ }\href {https://doi.org/10.1103/PhysRevD.86.025022} {\bibfield
  {journal} {\bibinfo  {journal} {Phys. Rev. D}\ }\textbf {\bibinfo {volume}
  {86}},\ \bibinfo {pages} {025022} (\bibinfo {year} {2012})},\ \Eprint
  {https://arxiv.org/abs/1203.6064} {arXiv:1203.6064 [hep-th]} \BibitemShut
  {NoStop}%
\bibitem [{\citenamefont {Poland}\ \emph {et~al.}(2019)\citenamefont {Poland},
  \citenamefont {Rychkov},\ and\ \citenamefont {Vichi}}]{Poland:2018epd}%
  \BibitemOpen
  \bibfield  {author} {\bibinfo {author} {\bibfnamefont {D.}~\bibnamefont
  {Poland}}, \bibinfo {author} {\bibfnamefont {S.}~\bibnamefont {Rychkov}},\
  and\ \bibinfo {author} {\bibfnamefont {A.}~\bibnamefont {Vichi}},\ }\bibfield
   {title} {\bibinfo {title} {{The Conformal Bootstrap: Theory, Numerical
  Techniques, and Applications}},\ }\href
  {https://doi.org/10.1103/RevModPhys.91.015002} {\bibfield  {journal}
  {\bibinfo  {journal} {Rev. Mod. Phys.}\ }\textbf {\bibinfo {volume} {91}},\
  \bibinfo {pages} {015002} (\bibinfo {year} {2019})},\ \Eprint
  {https://arxiv.org/abs/1805.04405} {arXiv:1805.04405 [hep-th]} \BibitemShut
  {NoStop}%
\bibitem [{\citenamefont {Han}\ \emph {et~al.}(2020)\citenamefont {Han},
  \citenamefont {Hartnoll},\ and\ \citenamefont {Kruthoff}}]{Han:2020bkb}%
  \BibitemOpen
  \bibfield  {author} {\bibinfo {author} {\bibfnamefont {X.}~\bibnamefont
  {Han}}, \bibinfo {author} {\bibfnamefont {S.~A.}\ \bibnamefont {Hartnoll}},\
  and\ \bibinfo {author} {\bibfnamefont {J.}~\bibnamefont {Kruthoff}},\
  }\bibfield  {title} {\bibinfo {title} {{Bootstrapping Matrix Quantum
  Mechanics}},\ }\href {https://doi.org/10.1103/PhysRevLett.125.041601}
  {\bibfield  {journal} {\bibinfo  {journal} {Phys. Rev. Lett.}\ }\textbf
  {\bibinfo {volume} {125}},\ \bibinfo {pages} {041601} (\bibinfo {year}
  {2020})},\ \Eprint {https://arxiv.org/abs/2004.10212} {arXiv:2004.10212
  [hep-th]} \BibitemShut {NoStop}%
\bibitem [{\citenamefont {Hirschfelder}(1960)}]{hirschfelder1960classical}%
  \BibitemOpen
  \bibfield  {author} {\bibinfo {author} {\bibfnamefont {J.~O.}\ \bibnamefont
  {Hirschfelder}},\ }\bibfield  {title} {\bibinfo {title} {Classical and
  quantum mechanical hypervirial theorems},\ }\href@noop {} {\bibfield
  {journal} {\bibinfo  {journal} {The Journal of Chemical Physics}\ }\textbf
  {\bibinfo {volume} {33}},\ \bibinfo {pages} {1462} (\bibinfo {year}
  {1960})}\BibitemShut {NoStop}%
\bibitem [{\citenamefont {Berenstein}\ and\ \citenamefont
  {Hulsey}(2021)}]{Berenstein:2021dyf}%
  \BibitemOpen
  \bibfield  {author} {\bibinfo {author} {\bibfnamefont {D.}~\bibnamefont
  {Berenstein}}\ and\ \bibinfo {author} {\bibfnamefont {G.}~\bibnamefont
  {Hulsey}},\ }\bibfield  {title} {\bibinfo {title} {{Bootstrapping Simple QM
  Systems}},\ }\href@noop {} {\  (\bibinfo {year} {2021})},\ \Eprint
  {https://arxiv.org/abs/2108.08757} {arXiv:2108.08757 [hep-th]} \BibitemShut
  {NoStop}%
\bibitem [{\citenamefont {Bhattacharya}\ \emph {et~al.}(2021)\citenamefont
  {Bhattacharya}, \citenamefont {Das}, \citenamefont {Das}, \citenamefont
  {Jha},\ and\ \citenamefont {Kundu}}]{Bhattacharya:2021btd}%
  \BibitemOpen
  \bibfield  {author} {\bibinfo {author} {\bibfnamefont {J.}~\bibnamefont
  {Bhattacharya}}, \bibinfo {author} {\bibfnamefont {D.}~\bibnamefont {Das}},
  \bibinfo {author} {\bibfnamefont {S.~K.}\ \bibnamefont {Das}}, \bibinfo
  {author} {\bibfnamefont {A.~K.}\ \bibnamefont {Jha}},\ and\ \bibinfo {author}
  {\bibfnamefont {M.}~\bibnamefont {Kundu}},\ }\bibfield  {title} {\bibinfo
  {title} {{Numerical bootstrap in quantum mechanics}},\ }\href
  {https://doi.org/10.1016/j.physletb.2021.136785} {\bibfield  {journal}
  {\bibinfo  {journal} {Phys. Lett. B}\ }\textbf {\bibinfo {volume} {823}},\
  \bibinfo {pages} {136785} (\bibinfo {year} {2021})},\ \Eprint
  {https://arxiv.org/abs/2108.11416} {arXiv:2108.11416 [hep-th]} \BibitemShut
  {NoStop}%
\bibitem [{\citenamefont {Aikawa}\ \emph {et~al.}(2022)\citenamefont {Aikawa},
  \citenamefont {Morita},\ and\ \citenamefont {Yoshimura}}]{Aikawa:2021eai}%
  \BibitemOpen
  \bibfield  {author} {\bibinfo {author} {\bibfnamefont {Y.}~\bibnamefont
  {Aikawa}}, \bibinfo {author} {\bibfnamefont {T.}~\bibnamefont {Morita}},\
  and\ \bibinfo {author} {\bibfnamefont {K.}~\bibnamefont {Yoshimura}},\
  }\bibfield  {title} {\bibinfo {title} {{Application of bootstrap to a
  {\ensuremath{\theta}} term}},\ }\href
  {https://doi.org/10.1103/PhysRevD.105.085017} {\bibfield  {journal} {\bibinfo
   {journal} {Phys. Rev. D}\ }\textbf {\bibinfo {volume} {105}},\ \bibinfo
  {pages} {085017} (\bibinfo {year} {2022})},\ \Eprint
  {https://arxiv.org/abs/2109.02701} {arXiv:2109.02701 [hep-th]} \BibitemShut
  {NoStop}%
\bibitem [{\citenamefont {Berenstein}\ and\ \citenamefont
  {Hulsey}(2022{\natexlab{a}})}]{Berenstein:2021loy}%
  \BibitemOpen
  \bibfield  {author} {\bibinfo {author} {\bibfnamefont {D.}~\bibnamefont
  {Berenstein}}\ and\ \bibinfo {author} {\bibfnamefont {G.}~\bibnamefont
  {Hulsey}},\ }\bibfield  {title} {\bibinfo {title} {{Bootstrapping more QM
  systems}},\ }\href {https://doi.org/10.1088/1751-8121/ac7118} {\bibfield
  {journal} {\bibinfo  {journal} {J. Phys. A}\ }\textbf {\bibinfo {volume}
  {55}},\ \bibinfo {pages} {275304} (\bibinfo {year} {2022}{\natexlab{a}})},\
  \Eprint {https://arxiv.org/abs/2109.06251} {arXiv:2109.06251 [hep-th]}
  \BibitemShut {NoStop}%
\bibitem [{\citenamefont {Tchoumakov}\ and\ \citenamefont
  {Florens}(2022)}]{Tchoumakov:2021mnh}%
  \BibitemOpen
  \bibfield  {author} {\bibinfo {author} {\bibfnamefont {S.}~\bibnamefont
  {Tchoumakov}}\ and\ \bibinfo {author} {\bibfnamefont {S.}~\bibnamefont
  {Florens}},\ }\bibfield  {title} {\bibinfo {title} {{Bootstrapping Bloch
  bands}},\ }\href {https://doi.org/10.1088/1751-8121/ac3c82} {\bibfield
  {journal} {\bibinfo  {journal} {J. Phys. A}\ }\textbf {\bibinfo {volume}
  {55}},\ \bibinfo {pages} {015203} (\bibinfo {year} {2022})},\ \Eprint
  {https://arxiv.org/abs/2109.06600} {arXiv:2109.06600 [cond-mat.mes-hall]}
  \BibitemShut {NoStop}%
\bibitem [{\citenamefont {Berenstein}\ and\ \citenamefont
  {Hulsey}(2023)}]{Berenstein:2022unr}%
  \BibitemOpen
  \bibfield  {author} {\bibinfo {author} {\bibfnamefont {D.}~\bibnamefont
  {Berenstein}}\ and\ \bibinfo {author} {\bibfnamefont {G.}~\bibnamefont
  {Hulsey}},\ }\bibfield  {title} {\bibinfo {title} {{Semidefinite programming
  algorithm for the quantum mechanical bootstrap}},\ }\href
  {https://doi.org/10.1103/PhysRevE.107.L053301} {\bibfield  {journal}
  {\bibinfo  {journal} {Phys. Rev. E}\ }\textbf {\bibinfo {volume} {107}},\
  \bibinfo {pages} {L053301} (\bibinfo {year} {2023})},\ \Eprint
  {https://arxiv.org/abs/2209.14332} {arXiv:2209.14332 [hep-th]} \BibitemShut
  {NoStop}%
\bibitem [{\citenamefont {Berenstein}\ and\ \citenamefont
  {Hulsey}(2022{\natexlab{b}})}]{Berenstein:2022ygg}%
  \BibitemOpen
  \bibfield  {author} {\bibinfo {author} {\bibfnamefont {D.}~\bibnamefont
  {Berenstein}}\ and\ \bibinfo {author} {\bibfnamefont {G.}~\bibnamefont
  {Hulsey}},\ }\bibfield  {title} {\bibinfo {title} {{Anomalous bootstrap on
  the half-line}},\ }\href {https://doi.org/10.1103/PhysRevD.106.045029}
  {\bibfield  {journal} {\bibinfo  {journal} {Phys. Rev. D}\ }\textbf {\bibinfo
  {volume} {106}},\ \bibinfo {pages} {045029} (\bibinfo {year}
  {2022}{\natexlab{b}})},\ \Eprint {https://arxiv.org/abs/2206.01765}
  {arXiv:2206.01765 [hep-th]} \BibitemShut {NoStop}%
\bibitem [{\citenamefont {Berenstein}\ and\ \citenamefont
  {Hulsey}(2024)}]{Berenstein:2023ppj}%
  \BibitemOpen
  \bibfield  {author} {\bibinfo {author} {\bibfnamefont {D.}~\bibnamefont
  {Berenstein}}\ and\ \bibinfo {author} {\bibfnamefont {G.}~\bibnamefont
  {Hulsey}},\ }\bibfield  {title} {\bibinfo {title} {{One-dimensional
  reflection in the quantum mechanical bootstrap}},\ }\href
  {https://doi.org/10.1103/PhysRevD.109.025013} {\bibfield  {journal} {\bibinfo
   {journal} {Phys. Rev. D}\ }\textbf {\bibinfo {volume} {109}},\ \bibinfo
  {pages} {025013} (\bibinfo {year} {2024})},\ \Eprint
  {https://arxiv.org/abs/2307.11724} {arXiv:2307.11724 [hep-th]} \BibitemShut
  {NoStop}%
\bibitem [{\citenamefont {Sword}\ and\ \citenamefont
  {Vegh}(2024)}]{Sword:2024gvv}%
  \BibitemOpen
  \bibfield  {author} {\bibinfo {author} {\bibfnamefont {L.}~\bibnamefont
  {Sword}}\ and\ \bibinfo {author} {\bibfnamefont {D.}~\bibnamefont {Vegh}},\
  }\bibfield  {title} {\bibinfo {title} {{Quantum mechanical bootstrap on the
  interval: Obtaining the exact spectrum}},\ }\href
  {https://doi.org/10.1103/PhysRevD.109.126002} {\bibfield  {journal} {\bibinfo
   {journal} {Phys. Rev. D}\ }\textbf {\bibinfo {volume} {109}},\ \bibinfo
  {pages} {126002} (\bibinfo {year} {2024})},\ \Eprint
  {https://arxiv.org/abs/2402.03434} {arXiv:2402.03434 [hep-th]} \BibitemShut
  {NoStop}%
\bibitem [{\citenamefont {Thong}\ and\ \citenamefont
  {Vegh}(2026)}]{Thong:2026zvt}%
  \BibitemOpen
  \bibfield  {author} {\bibinfo {author} {\bibfnamefont {K.~H.}\ \bibnamefont
  {Thong}}\ and\ \bibinfo {author} {\bibfnamefont {D.}~\bibnamefont {Vegh}},\
  }\bibfield  {title} {\bibinfo {title} {{Quantum mechanical bootstrap without
  inequalities: SYK bilinear spectrum}},\ }\href@noop {} {\  (\bibinfo {year}
  {2026})},\ \Eprint {https://arxiv.org/abs/2604.26007} {arXiv:2604.26007
  [hep-th]} \BibitemShut {NoStop}%
\bibitem [{\citenamefont {Berenstein}\ and\ \citenamefont
  {Rodriguez}(2025)}]{Berenstein:2025itw}%
  \BibitemOpen
  \bibfield  {author} {\bibinfo {author} {\bibfnamefont {D.}~\bibnamefont
  {Berenstein}}\ and\ \bibinfo {author} {\bibfnamefont {V.~A.}\ \bibnamefont
  {Rodriguez}},\ }\bibfield  {title} {\bibinfo {title} {{Goldilocks and the
  bootstrap}},\ }\href {https://doi.org/10.1007/JHEP09(2025)109} {\bibfield
  {journal} {\bibinfo  {journal} {JHEP}\ }\textbf {\bibinfo {volume} {09}},\
  \bibinfo {pages} {109}},\ \Eprint {https://arxiv.org/abs/2503.00104}
  {arXiv:2503.00104 [hep-th]} \BibitemShut {NoStop}%
\bibitem [{\citenamefont {Berenstein}\ \emph {et~al.}(2026)\citenamefont
  {Berenstein}, \citenamefont {Rodrigues},\ and\ \citenamefont
  {Rodriguez}}]{Berenstein:2026wky}%
  \BibitemOpen
  \bibfield  {author} {\bibinfo {author} {\bibfnamefont {D.}~\bibnamefont
  {Berenstein}}, \bibinfo {author} {\bibfnamefont {J.}~\bibnamefont
  {Rodrigues}},\ and\ \bibinfo {author} {\bibfnamefont {V.~A.}\ \bibnamefont
  {Rodriguez}},\ }\bibfield  {title} {\bibinfo {title} {{Asymptotic bootstrap
  for unitary matrix integrals at complex coupling}},\ }\href@noop {} {\
  (\bibinfo {year} {2026})},\ \Eprint {https://arxiv.org/abs/2602.18559}
  {arXiv:2602.18559 [hep-th]} \BibitemShut {NoStop}%
\bibitem [{\citenamefont {Schm{\"u}dgen}(2017)}]{schmudgen2017moment}%
  \BibitemOpen
  \bibfield  {author} {\bibinfo {author} {\bibfnamefont {K.}~\bibnamefont
  {Schm{\"u}dgen}},\ }\href@noop {} {\emph {\bibinfo {title} {The moment
  problem}}},\ Vol.~\bibinfo {volume} {9}\ (\bibinfo  {publisher} {Springer},\
  \bibinfo {year} {2017})\BibitemShut {NoStop}%
\bibitem [{\citenamefont {Fitzpatrick}\ \emph {et~al.}(2013)\citenamefont
  {Fitzpatrick}, \citenamefont {Kaplan}, \citenamefont {Poland},\ and\
  \citenamefont {Simmons-Duffin}}]{Fitzpatrick:2012yx}%
  \BibitemOpen
  \bibfield  {author} {\bibinfo {author} {\bibfnamefont {A.~L.}\ \bibnamefont
  {Fitzpatrick}}, \bibinfo {author} {\bibfnamefont {J.}~\bibnamefont {Kaplan}},
  \bibinfo {author} {\bibfnamefont {D.}~\bibnamefont {Poland}},\ and\ \bibinfo
  {author} {\bibfnamefont {D.}~\bibnamefont {Simmons-Duffin}},\ }\bibfield
  {title} {\bibinfo {title} {{The Analytic Bootstrap and AdS Superhorizon
  Locality}},\ }\href {https://doi.org/10.1007/JHEP12(2013)004} {\bibfield
  {journal} {\bibinfo  {journal} {JHEP}\ }\textbf {\bibinfo {volume} {12}},\
  \bibinfo {pages} {004}},\ \Eprint {https://arxiv.org/abs/1212.3616}
  {arXiv:1212.3616 [hep-th]} \BibitemShut {NoStop}%
\bibitem [{\citenamefont {Hartman}\ \emph {et~al.}(2022)\citenamefont
  {Hartman}, \citenamefont {Mazac}, \citenamefont {Simmons-Duffin},\ and\
  \citenamefont {Zhiboedov}}]{Hartman:2022zik}%
  \BibitemOpen
  \bibfield  {author} {\bibinfo {author} {\bibfnamefont {T.}~\bibnamefont
  {Hartman}}, \bibinfo {author} {\bibfnamefont {D.}~\bibnamefont {Mazac}},
  \bibinfo {author} {\bibfnamefont {D.}~\bibnamefont {Simmons-Duffin}},\ and\
  \bibinfo {author} {\bibfnamefont {A.}~\bibnamefont {Zhiboedov}},\ }\bibfield
  {title} {\bibinfo {title} {{Snowmass White Paper: The Analytic Conformal
  Bootstrap}},\ }in\ \href@noop {} {\emph {\bibinfo {booktitle} {{Snowmass
  2021}}}}\ (\bibinfo {year} {2022})\ \Eprint
  {https://arxiv.org/abs/2202.11012} {arXiv:2202.11012 [hep-th]} \BibitemShut
  {NoStop}%
\bibitem [{\citenamefont {Huang}\ and\ \citenamefont
  {Li}(2025)}]{Huang:2025sua}%
  \BibitemOpen
  \bibfield  {author} {\bibinfo {author} {\bibfnamefont {Z.}~\bibnamefont
  {Huang}}\ and\ \bibinfo {author} {\bibfnamefont {W.}~\bibnamefont {Li}},\
  }\bibfield  {title} {\bibinfo {title} {{Bootstrapping periodic quantum
  systems}},\ }\href@noop {} {\  (\bibinfo {year} {2025})},\ \Eprint
  {https://arxiv.org/abs/2507.02386} {arXiv:2507.02386 [hep-th]} \BibitemShut
  {NoStop}%
\bibitem [{\citenamefont {Fiol}\ \emph {et~al.}(2026)\citenamefont {Fiol},
  \citenamefont {Gijon},\ and\ \citenamefont {Alonso}}]{Fiol:2025gkq}%
  \BibitemOpen
  \bibfield  {author} {\bibinfo {author} {\bibfnamefont {B.}~\bibnamefont
  {Fiol}}, \bibinfo {author} {\bibfnamefont {E.}~\bibnamefont {Gijon}},\ and\
  \bibinfo {author} {\bibfnamefont {U.~L.}\ \bibnamefont {Alonso}},\ }\bibfield
   {title} {\bibinfo {title} {{Schwinger-Dyson approximants}},\ }\href
  {https://doi.org/10.1007/JHEP02(2026)111} {\bibfield  {journal} {\bibinfo
  {journal} {JHEP}\ }\textbf {\bibinfo {volume} {02}},\ \bibinfo {pages}
  {111}},\ \Eprint {https://arxiv.org/abs/2511.05665} {arXiv:2511.05665
  [hep-th]} \BibitemShut {NoStop}%
\bibitem [{\citenamefont {Wigner}(1993)}]{wigner1993characteristic}%
  \BibitemOpen
  \bibfield  {author} {\bibinfo {author} {\bibfnamefont {E.~P.}\ \bibnamefont
  {Wigner}},\ }\bibfield  {title} {\bibinfo {title} {Characteristic vectors of
  bordered matrices with infinite dimensions i},\ }in\ \href@noop {} {\emph
  {\bibinfo {booktitle} {The Collected Works of Eugene Paul Wigner: Part A: The
  Scientific Papers}}}\ (\bibinfo  {publisher} {Springer},\ \bibinfo {year}
  {1993})\ pp.\ \bibinfo {pages} {524--540}\BibitemShut {NoStop}%
\bibitem [{\citenamefont {Dyson}(1962)}]{Dyson:1962es}%
  \BibitemOpen
  \bibfield  {author} {\bibinfo {author} {\bibfnamefont {F.~J.}\ \bibnamefont
  {Dyson}},\ }\bibfield  {title} {\bibinfo {title} {{Statistical theory of the
  energy levels of complex systems. I}},\ }\href
  {https://doi.org/10.1063/1.1703773} {\bibfield  {journal} {\bibinfo
  {journal} {J. Math. Phys.}\ }\textbf {\bibinfo {volume} {3}},\ \bibinfo
  {pages} {140} (\bibinfo {year} {1962})}\BibitemShut {NoStop}%
\bibitem [{\citenamefont {Dyson}(1970)}]{Dyson:1970tza}%
  \BibitemOpen
  \bibfield  {author} {\bibinfo {author} {\bibfnamefont {F.~J.}\ \bibnamefont
  {Dyson}},\ }\bibfield  {title} {\bibinfo {title} {{Correlations between the
  eigenvalues of a random matrix}},\ }\href
  {https://doi.org/10.1007/BF01646824} {\bibfield  {journal} {\bibinfo
  {journal} {Commun. Math. Phys.}\ }\textbf {\bibinfo {volume} {19}},\ \bibinfo
  {pages} {235} (\bibinfo {year} {1970})}\BibitemShut {NoStop}%
\bibitem [{\citenamefont {D'Alessio}\ \emph {et~al.}(2016)\citenamefont
  {D'Alessio}, \citenamefont {Kafri}, \citenamefont {Polkovnikov},\ and\
  \citenamefont {Rigol}}]{DAlessio:2015qtq}%
  \BibitemOpen
  \bibfield  {author} {\bibinfo {author} {\bibfnamefont {L.}~\bibnamefont
  {D'Alessio}}, \bibinfo {author} {\bibfnamefont {Y.}~\bibnamefont {Kafri}},
  \bibinfo {author} {\bibfnamefont {A.}~\bibnamefont {Polkovnikov}},\ and\
  \bibinfo {author} {\bibfnamefont {M.}~\bibnamefont {Rigol}},\ }\bibfield
  {title} {\bibinfo {title} {{From quantum chaos and eigenstate thermalization
  to statistical mechanics and thermodynamics}},\ }\href
  {https://doi.org/10.1080/00018732.2016.1198134} {\bibfield  {journal}
  {\bibinfo  {journal} {Adv. Phys.}\ }\textbf {\bibinfo {volume} {65}},\
  \bibinfo {pages} {239} (\bibinfo {year} {2016})},\ \Eprint
  {https://arxiv.org/abs/1509.06411} {arXiv:1509.06411 [cond-mat.stat-mech]}
  \BibitemShut {NoStop}%
\bibitem [{\citenamefont {Saad}\ \emph {et~al.}(2019)\citenamefont {Saad},
  \citenamefont {Shenker},\ and\ \citenamefont {Stanford}}]{Saad:2019lba}%
  \BibitemOpen
  \bibfield  {author} {\bibinfo {author} {\bibfnamefont {P.}~\bibnamefont
  {Saad}}, \bibinfo {author} {\bibfnamefont {S.~H.}\ \bibnamefont {Shenker}},\
  and\ \bibinfo {author} {\bibfnamefont {D.}~\bibnamefont {Stanford}},\
  }\bibfield  {title} {\bibinfo {title} {{JT gravity as a matrix integral}},\
  }\href@noop {} {\  (\bibinfo {year} {2019})},\ \Eprint
  {https://arxiv.org/abs/1903.11115} {arXiv:1903.11115 [hep-th]} \BibitemShut
  {NoStop}%
\bibitem [{\citenamefont {Dijkgraaf}\ and\ \citenamefont
  {Vafa}(2002)}]{Dijkgraaf:2002vw}%
  \BibitemOpen
  \bibfield  {author} {\bibinfo {author} {\bibfnamefont {R.}~\bibnamefont
  {Dijkgraaf}}\ and\ \bibinfo {author} {\bibfnamefont {C.}~\bibnamefont
  {Vafa}},\ }\bibfield  {title} {\bibinfo {title} {{On geometry and matrix
  models}},\ }\href {https://doi.org/10.1016/S0550-3213(02)00764-2} {\bibfield
  {journal} {\bibinfo  {journal} {Nucl. Phys. B}\ }\textbf {\bibinfo {volume}
  {644}},\ \bibinfo {pages} {21} (\bibinfo {year} {2002})},\ \Eprint
  {https://arxiv.org/abs/hep-th/0207106} {arXiv:hep-th/0207106} \BibitemShut
  {NoStop}%
\bibitem [{\citenamefont {Lin}(2020)}]{Lin:2020mme}%
  \BibitemOpen
  \bibfield  {author} {\bibinfo {author} {\bibfnamefont {H.~W.}\ \bibnamefont
  {Lin}},\ }\bibfield  {title} {\bibinfo {title} {{Bootstraps to strings:
  solving random matrix models with positivity}},\ }\href
  {https://doi.org/10.1007/JHEP06(2020)090} {\bibfield  {journal} {\bibinfo
  {journal} {JHEP}\ }\textbf {\bibinfo {volume} {06}},\ \bibinfo {pages}
  {090}},\ \Eprint {https://arxiv.org/abs/2002.08387} {arXiv:2002.08387
  [hep-th]} \BibitemShut {NoStop}%
\bibitem [{\citenamefont {Anderson}\ and\ \citenamefont
  {Kruczenski}(2017)}]{Anderson:2016rcw}%
  \BibitemOpen
  \bibfield  {author} {\bibinfo {author} {\bibfnamefont {P.~D.}\ \bibnamefont
  {Anderson}}\ and\ \bibinfo {author} {\bibfnamefont {M.}~\bibnamefont
  {Kruczenski}},\ }\bibfield  {title} {\bibinfo {title} {{Loop Equations and
  bootstrap methods in the lattice}},\ }\href
  {https://doi.org/10.1016/j.nuclphysb.2017.06.009} {\bibfield  {journal}
  {\bibinfo  {journal} {Nucl. Phys. B}\ }\textbf {\bibinfo {volume} {921}},\
  \bibinfo {pages} {702} (\bibinfo {year} {2017})},\ \Eprint
  {https://arxiv.org/abs/1612.08140} {arXiv:1612.08140 [hep-th]} \BibitemShut
  {NoStop}%
\bibitem [{\citenamefont {Brezin}\ \emph {et~al.}(1978)\citenamefont {Brezin},
  \citenamefont {Itzykson}, \citenamefont {Parisi},\ and\ \citenamefont
  {Zuber}}]{Brezin:1977sv}%
  \BibitemOpen
  \bibfield  {author} {\bibinfo {author} {\bibfnamefont {E.}~\bibnamefont
  {Brezin}}, \bibinfo {author} {\bibfnamefont {C.}~\bibnamefont {Itzykson}},
  \bibinfo {author} {\bibfnamefont {G.}~\bibnamefont {Parisi}},\ and\ \bibinfo
  {author} {\bibfnamefont {J.~B.}\ \bibnamefont {Zuber}},\ }\bibfield  {title}
  {\bibinfo {title} {{Planar Diagrams}},\ }\href
  {https://doi.org/10.1007/BF01614153} {\bibfield  {journal} {\bibinfo
  {journal} {Commun. Math. Phys.}\ }\textbf {\bibinfo {volume} {59}},\ \bibinfo
  {pages} {35} (\bibinfo {year} {1978})}\BibitemShut {NoStop}%
\bibitem [{\citenamefont {Kazakov}\ and\ \citenamefont
  {Zheng}(2022)}]{Kazakov:2021lel}%
  \BibitemOpen
  \bibfield  {author} {\bibinfo {author} {\bibfnamefont {V.}~\bibnamefont
  {Kazakov}}\ and\ \bibinfo {author} {\bibfnamefont {Z.}~\bibnamefont
  {Zheng}},\ }\bibfield  {title} {\bibinfo {title} {{Analytic and numerical
  bootstrap for one-matrix model and
  {\textquotedblleft}unsolvable{\textquotedblright} two-matrix model}},\ }\href
  {https://doi.org/10.1007/JHEP06(2022)030} {\bibfield  {journal} {\bibinfo
  {journal} {JHEP}\ }\textbf {\bibinfo {volume} {06}},\ \bibinfo {pages}
  {030}},\ \Eprint {https://arxiv.org/abs/2108.04830} {arXiv:2108.04830
  [hep-th]} \BibitemShut {NoStop}%
\bibitem [{\citenamefont {Kov{\'a}{\v{c}}ik}\ and\ \citenamefont
  {Magdolenov{\'a}}(2025)}]{Kovacik:2025qgj}%
  \BibitemOpen
  \bibfield  {author} {\bibinfo {author} {\bibfnamefont {S.}~\bibnamefont
  {Kov{\'a}{\v{c}}ik}}\ and\ \bibinfo {author} {\bibfnamefont {K.}~\bibnamefont
  {Magdolenov{\'a}}},\ }\bibfield  {title} {\bibinfo {title} {{Eigenvalue
  distribution from bootstrap estimates}},\ }\href
  {https://doi.org/10.1103/m812-s3hr} {\bibfield  {journal} {\bibinfo
  {journal} {Phys. Rev. D}\ }\textbf {\bibinfo {volume} {112}},\ \bibinfo
  {pages} {126021} (\bibinfo {year} {2025})},\ \Eprint
  {https://arxiv.org/abs/2509.16005} {arXiv:2509.16005 [hep-th]} \BibitemShut
  {NoStop}%
\bibitem [{\citenamefont {Bessis}(1979)}]{Bessis:1979is}%
  \BibitemOpen
  \bibfield  {author} {\bibinfo {author} {\bibfnamefont {D.}~\bibnamefont
  {Bessis}},\ }\bibfield  {title} {\bibinfo {title} {{A New Method in the
  Combinatorics of the Topological Expansion}},\ }\href
  {https://doi.org/10.1007/BF01221445} {\bibfield  {journal} {\bibinfo
  {journal} {Commun. Math. Phys.}\ }\textbf {\bibinfo {volume} {69}},\ \bibinfo
  {pages} {147} (\bibinfo {year} {1979})}\BibitemShut {NoStop}%
\bibitem [{\citenamefont {Itzykson}\ and\ \citenamefont
  {Zuber}(1980)}]{Itzykson:1979fi}%
  \BibitemOpen
  \bibfield  {author} {\bibinfo {author} {\bibfnamefont {C.}~\bibnamefont
  {Itzykson}}\ and\ \bibinfo {author} {\bibfnamefont {J.~B.}\ \bibnamefont
  {Zuber}},\ }\bibfield  {title} {\bibinfo {title} {{The Planar Approximation.
  2.}},\ }\href {https://doi.org/10.1063/1.524438} {\bibfield  {journal}
  {\bibinfo  {journal} {J. Math. Phys.}\ }\textbf {\bibinfo {volume} {21}},\
  \bibinfo {pages} {411} (\bibinfo {year} {1980})}\BibitemShut {NoStop}%
\bibitem [{\citenamefont {'t~Hooft}(1974)}]{tHooft:1973alw}%
  \BibitemOpen
  \bibfield  {author} {\bibinfo {author} {\bibfnamefont {G.}~\bibnamefont
  {'t~Hooft}},\ }\bibfield  {title} {\bibinfo {title} {{A Planar Diagram Theory
  for Strong Interactions}},\ }\href
  {https://doi.org/10.1016/0550-3213(74)90154-0} {\bibfield  {journal}
  {\bibinfo  {journal} {Nucl. Phys. B}\ }\textbf {\bibinfo {volume} {72}},\
  \bibinfo {pages} {461} (\bibinfo {year} {1974})}\BibitemShut {NoStop}%
\bibitem [{\citenamefont {Gautschi}(1982)}]{gautschi1982generating}%
  \BibitemOpen
  \bibfield  {author} {\bibinfo {author} {\bibfnamefont {W.}~\bibnamefont
  {Gautschi}},\ }\bibfield  {title} {\bibinfo {title} {On generating orthogonal
  polynomials},\ }\href@noop {} {\bibfield  {journal} {\bibinfo  {journal}
  {SIAM Journal on Scientific and Statistical Computing}\ }\textbf {\bibinfo
  {volume} {3}},\ \bibinfo {pages} {289} (\bibinfo {year} {1982})}\BibitemShut
  {NoStop}%
\bibitem [{\citenamefont {Gautschi}(1985)}]{gautschi1985orthogonal}%
  \BibitemOpen
  \bibfield  {author} {\bibinfo {author} {\bibfnamefont {W.}~\bibnamefont
  {Gautschi}},\ }\bibfield  {title} {\bibinfo {title} {Orthogonal
  polynomials—constructive theory and applications},\ }\href@noop {}
  {\bibfield  {journal} {\bibinfo  {journal} {Journal of Computational and
  Applied Mathematics}\ }\textbf {\bibinfo {volume} {12}},\ \bibinfo {pages}
  {61} (\bibinfo {year} {1985})}\BibitemShut {NoStop}%
\bibitem [{\citenamefont {Gautschi}(2004)}]{gautschi2004orthogonal}%
  \BibitemOpen
  \bibfield  {author} {\bibinfo {author} {\bibfnamefont {W.}~\bibnamefont
  {Gautschi}},\ }\href@noop {} {\emph {\bibinfo {title} {Orthogonal
  polynomials: computation and approximation}}}\ (\bibinfo  {publisher} {OUP
  Oxford},\ \bibinfo {year} {2004})\BibitemShut {NoStop}%
\bibitem [{\citenamefont {Johnson}(2022)}]{Johnson:2022wsr}%
  \BibitemOpen
  \bibfield  {author} {\bibinfo {author} {\bibfnamefont {C.~V.}\ \bibnamefont
  {Johnson}},\ }\bibfield  {title} {\bibinfo {title} {{The Microstate Physics
  of JT Gravity and Supergravity}},\ }\href@noop {} {\  (\bibinfo {year}
  {2022})},\ \Eprint {https://arxiv.org/abs/2201.11942} {arXiv:2201.11942
  [hep-th]} \BibitemShut {NoStop}%
\bibitem [{\citenamefont {Mari{\~n}o}(2014)}]{Marino:2012zq}%
  \BibitemOpen
  \bibfield  {author} {\bibinfo {author} {\bibfnamefont {M.}~\bibnamefont
  {Mari{\~n}o}},\ }\bibfield  {title} {\bibinfo {title} {{Lectures on
  non-perturbative effects in large $N$ gauge theories, matrix models and
  strings}},\ }\href {https://doi.org/10.1002/prop.201400005} {\bibfield
  {journal} {\bibinfo  {journal} {Fortsch. Phys.}\ }\textbf {\bibinfo {volume}
  {62}},\ \bibinfo {pages} {455} (\bibinfo {year} {2014})},\ \Eprint
  {https://arxiv.org/abs/1206.6272} {arXiv:1206.6272 [hep-th]} \BibitemShut
  {NoStop}%
\bibitem [{\citenamefont {Kager}\ \emph {et~al.}(2026)\citenamefont {Kager},
  \citenamefont {Rodrigues}, \citenamefont {Schiappa}, \citenamefont
  {Schwick},\ and\ \citenamefont {Tamarin}}]{Kager:2026cod}%
  \BibitemOpen
  \bibfield  {author} {\bibinfo {author} {\bibfnamefont {J.}~\bibnamefont
  {Kager}}, \bibinfo {author} {\bibfnamefont {J.}~\bibnamefont {Rodrigues}},
  \bibinfo {author} {\bibfnamefont {R.}~\bibnamefont {Schiappa}}, \bibinfo
  {author} {\bibfnamefont {M.}~\bibnamefont {Schwick}},\ and\ \bibinfo {author}
  {\bibfnamefont {N.}~\bibnamefont {Tamarin}},\ }\bibfield  {title} {\bibinfo
  {title} {{Exact Solutions to Matrix Models and String Theories: The Local
  Construction}},\ }\href@noop {} {\  (\bibinfo {year} {2026})},\ \Eprint
  {https://arxiv.org/abs/2602.15101} {arXiv:2602.15101 [hep-th]} \BibitemShut
  {NoStop}%
\bibitem [{\citenamefont {Anninos}\ and\ \citenamefont
  {M{\"u}hlmann}(2020)}]{Anninos:2020ccj}%
  \BibitemOpen
  \bibfield  {author} {\bibinfo {author} {\bibfnamefont {D.}~\bibnamefont
  {Anninos}}\ and\ \bibinfo {author} {\bibfnamefont {B.}~\bibnamefont
  {M{\"u}hlmann}},\ }\bibfield  {title} {\bibinfo {title} {{Notes on matrix
  models (matrix musings)}},\ }\href {https://doi.org/10.1088/1742-5468/aba499}
  {\bibfield  {journal} {\bibinfo  {journal} {J. Stat. Mech.}\ }\textbf
  {\bibinfo {volume} {2008}},\ \bibinfo {pages} {083109} (\bibinfo {year}
  {2020})},\ \Eprint {https://arxiv.org/abs/2004.01171} {arXiv:2004.01171
  [hep-th]} \BibitemShut {NoStop}%
\bibitem [{\citenamefont {Cachazo}\ \emph {et~al.}(2002)\citenamefont
  {Cachazo}, \citenamefont {Douglas}, \citenamefont {Seiberg},\ and\
  \citenamefont {Witten}}]{Cachazo:2002ry}%
  \BibitemOpen
  \bibfield  {author} {\bibinfo {author} {\bibfnamefont {F.}~\bibnamefont
  {Cachazo}}, \bibinfo {author} {\bibfnamefont {M.~R.}\ \bibnamefont
  {Douglas}}, \bibinfo {author} {\bibfnamefont {N.}~\bibnamefont {Seiberg}},\
  and\ \bibinfo {author} {\bibfnamefont {E.}~\bibnamefont {Witten}},\
  }\bibfield  {title} {\bibinfo {title} {{Chiral rings and anomalies in
  supersymmetric gauge theory}},\ }\href
  {https://doi.org/10.1088/1126-6708/2002/12/071} {\bibfield  {journal}
  {\bibinfo  {journal} {JHEP}\ }\textbf {\bibinfo {volume} {12}},\ \bibinfo
  {pages} {071}},\ \Eprint {https://arxiv.org/abs/hep-th/0211170}
  {arXiv:hep-th/0211170} \BibitemShut {NoStop}%
\bibitem [{\citenamefont {Marino}(2004)}]{Marino:2004eq}%
  \BibitemOpen
  \bibfield  {author} {\bibinfo {author} {\bibfnamefont {M.}~\bibnamefont
  {Marino}},\ }\bibfield  {title} {\bibinfo {title} {{Les Houches lectures on
  matrix models and topological strings}}\ }(\bibinfo {year} {2004})\ \Eprint
  {https://arxiv.org/abs/hep-th/0410165} {arXiv:hep-th/0410165} \BibitemShut
  {NoStop}%
\end{thebibliography}%

\end{document}